\documentclass[aps,prd,groupedaddress,showpacs,showkeys]{revtex4}
\usepackage{amssymb}

\begin{document}

\title{Scalar nonet quarkonia and the scalar glueball: \\
mixing and decays in an effective chiral approach}
\author{F. Giacosa, Th. Gutsche, V. E. Lyubovitskij and Amand Faessler}
\affiliation{Institut f\"ur Theoretische Physik, Universit\"at T\"ubingen, \\
Auf der Morgenstelle 14,D-72076 T\"ubingen, Germany\\}

\date{\today}

\begin{abstract}
We study the strong and electromagnetic decay properties of scalar mesons
above 1~GeV within a chiral approach. The scalar-isoscalar states are
treated as mixed states of quarkonia and glueball configurations. A fit to
the experimental mass and decay rates listed by the Particle Data group is
performed to extract phenomenological constraints on the nature of the
scalar resonances and to the issue of the glueball decays. A comparison to
other experimental results and to other theoretical approaches in the scalar
meson sector is discussed.
\end{abstract}

\pacs{12.39.Fe, 12.39.Mk, 13.25.Jx, 14.40.Cs}
\keywords{Scalar and pseudoscalar mesons, glueball, effective chiral
approach, strong and electromagnetic decays}
\maketitle

\vskip .5cm

\vskip .5cm

\setlength{\baselineskip}{1.5\baselineskip}

\section{Introduction}

\label{intro}

The unique interpretation of scalar mesons constitutes an unsolved problem
of hadronic QCD. Below the mass scale of 2~GeV various scalar states~\cite%
{Eidelman:2004wy} are encountered: the isoscalar resonances $\sigma =
f_{0}(400-1200)$, $f_{0}(980)$, $f_{0}(1370)$, $f_{0}(1500)$ and $%
f_{0}(1710) $, the isovectors $a_{0}(980)$ and $a_{0}(1450)$ and the
isodoublets $K_{0}^{\ast }(800)$ and $K_{0}^{\ast }(1430).$ The existence of
the $K_{0}^{\ast }(800)$ is still not well established and omitted from the
summary tables of~\cite{Eidelman:2004wy}. From a theoretical point of view
one expects the scalar quark-antiquark ground state nonet $0^{++}$, a
scalar-isoscalar glueball, which lattice QCD predicts to be the lightest
gluonic meson with a mass between 1.4-1.8 GeV~\cite{Michael:2003ai}, and
possibly other exotic states (non $\bar{q}q$ states), like e.g. four quark
states or mesonic molecules~\cite{Amsler:2004ps}. Various interpretations of
and assignments for the physical scalar resonances in terms of the expected
theoretical states have been proposed (see, for instance, the review 
papers~\cite{Amsler:2004ps,Close:2002zu,Narison:1996fm} and references therein).

In this work we follow the original assignment of Ref.~\cite{Amsler:1995tu},
where in a minimal scenario the bare quarkonia states 
$N\equiv \sqrt{1/2}(\bar{u}u+\bar{d}d)=\bar{n}n$, $S\equiv \bar{s}s$ 
and the bare scalar
glueball $G$ mix, resulting in the three scalar-isoscalar resonances 
$f_{0}(1370)$, $f_{0}(1500)$ and $f_{0}(1710)$. Such a mixing scheme has been
previously investigated by many authors, as for example in the lattice study
of~\cite{Lee:1999kv} or within the model approaches 
of~\cite{Close:2001ga}-\cite{Giacosa:2005qr}.

The mesons $a_{0}(1450)$ and $K_{0}^{\ast }(1430)$ are considered as the 
$I=1 $ and $I=1/2$ quarkonia $J^{PC}=0^{++}$ states. In this way the
low-lying scalar quarkonia nonet is located in the energy range of 1-2 GeV,
where the other p-wave nonets of tensor ($2^{++}$) and pseudovector mesons 
($1^{++}$) \cite{Eidelman:2004wy} are also situated. Masses of selected scalar
quarkonia states were also estimated on the Lattice; 
in Ref.~\cite{Prelovsek:2004jp} the $I=1$ scalar quarkonium state is 
predicted to have a mass of $M_{a_{0}}=1.51\pm 0.19$. 
This result favours the interpretation of
the state $a_{0}(1450)$ (and not $a_{0}(980)$) as the isovector ground-state
scalar quarkonium. However, previous lattice studies 
(as in~\cite{McNeile:2000xx}; see also \cite{Prelovsek:2004jp,Bali:2003tp} 
and Refs. therein) find different results. Scalar resonances below 1 GeV 
can possibly be interpreted as four quark states or mesonic molecules. 
In~\cite{Alford:2000mm} the hypothesis of Jaffe's four quark states was studied 
on the Lattice, where the masses of four-quark states are found to be lighter
than 1 GeV. Recent attempts in the context of chiral perturbation theory to
describe the scalar states below $1$ GeV as \textquotedblright dynamically
generated\textquotedblright\ resonances, i.e. states which do not survive in
the large $N_{c}$ limit \cite{Pelaez:2003dy,Oller:1997ti}, have also been
performed. In this scheme the nature of the scalars below $1$ GeV can also
be related to four quark configurations. This can be viewed as a further
indication that scalar quarkonia states are located at masses above 1~GeV.

In this work, starting from 
an effective chiral Lagrangian~\cite{Giacosa:2005qr} derived in
Chiral Perturbation Theory (ChPT)~\cite{Cirigliano:2003yq,ChPT,Ecker:1988te},  
we perform a tree-level analysis of the strong and
electromagnetic decays of scalar mesons settled in the energy range between
1 and 2 GeV. The scalar glueball is introduced as an extra-flavor singlet
composite field with independent couplings to pseudoscalar mesons (and to
photons, although suppressed). Although a chiral approach cannot be
rigorously justified at this energy scale, since loop corrections could be
large, we intend to use this framework as a phenomenological tool to extract
possible glueball-quarkonia mixing scenarios from the observed decays.

The scalar glueball $G$ mixes with the scalar quarkonia fields $N \equiv 
\sqrt{1/2}(\overline{u}u+\overline{d}d) = \overline{n} n$ and $S \equiv 
\overline{s}s$ in accord with flavor blindness. We also consider in this
scheme a possible direct mixing of the quarkonia fields $N$ and \ $S$. The
origin of such mixing can be driven by instantons~\cite{Klempt:1995ku}. The
presence of a (even small) flavor mixing in the scalar-isoscalar sector can
sensibly affect the phenomenology.

In the presented approach the glueball decay into two pseudoscalar mesons is
occurring by two mechanisms: a)~through mixing, that is the glueball $G$
acquires a quarkonium component, which subsequently decays into two
pseudoscalars; b)~direct decay of the glueball component $G$ into two
pseudoscalars without an intermediate scalar 
quarkonium~\cite{Amsler:1995tu,Close:2001ga,Strohmeier-Presicek:1999yv}. 
In~\cite{Amsler:1995tu,Strohmeier-Presicek:1999yv} 
this direct decay is argued to be
suppressed as based on arguments of the strong coupling expansion, while in
the phenomenological fit of~\cite{Close:2001ga} it dominates. Here we also
address the question, if the direct decay is needed to explain the decay
phenomenology of the scalar-isoscalar resonances $f_{0}(1370),$ $f_{0}(1500)$
and $f_{0}(1710)$.

For consistency we also consider the strong decays of the isovector and
isodoublet scalar states as well, trying to highlight the difficulties and
the open issues, also comparing with previous works on this subject.
Following the idea of~\cite{Cirigliano:2003yq} we analyze deviations from
the large $N_{c}$ limit in the framework of the proposed scalar-quarkonia
assignment (which differs from~\cite{Cirigliano:2003yq}), which turn out to
be relatively small in the phenomenology.

The paper is organized as follows: in Section~\ref{model} we discuss the
effective Lagrangian related to the scalar quarkonia-glueball mass spectrum
and the strong and electromagnetic decays. In Section~\ref{fit} we determine
a phenomenological fit to the experimental data listed in~\cite%
{Eidelman:2004wy} by first neglecting the direct decay of the glueball
component. In Section~\ref{phen} we also allow for a direct glueball decay
studying its influence on the results. There we use the lattice data 
of \cite{Sexton:1996ed}, where an approximate calculation of the glueball 
decays into two pseudoscalar mesons has been performed, 
to constrain our analysis. Finally, in Section~\ref{concl} 
we summarize our results and draw conclusions.

\section{The model}

\label{model}

\subsection{The Lagrangian}

The strong and electromagnetic decays of scalar mesons are based on an
effective chiral Lagrangian $\mathcal{L}_{\mathrm{eff}}$ as derived in
Chiral Perturbation Theory (ChPT)~\cite{Cirigliano:2003yq,ChPT,Ecker:1988te}. 
The Lagrangian involves the nonets of pseudoscalar and of scalar mesons, 
\begin{equation}
\mathcal{P}=\frac{1}{\sqrt{2}}\sum_{i=0}^{8}P_{i}\lambda _{i}\,,\hspace*{1cm} 
\mathcal{S}=\frac{1}{\sqrt{2}}\sum_{i=0}^{8}S_{i}\lambda _{i},  \label{ps}
\end{equation}%
the electromagnetic field and, in addition, a new degree of freedom, the
bare glueball field $G$, which is treated as a flavor-blind mesonic field.
The lowest order effective Lagrangian $\mathcal{L}_{\mathrm{eff}}$ in the
large $N_{c}$ limit including $1/N_{c}$ corrections reads 
\begin{eqnarray}
\mathcal{L}_{\mathrm{eff}} &=&\frac{F^{2}}{4}\left\langle D_{\mu }U\,D^{\mu
}U^{\dagger }+\chi _{+}\right\rangle +\frac{1}{2}\left\langle D_{\mu } 
\mathcal{S}D^{\mu }\mathcal{S}-M_{\mathcal{S}}^{2}\mathcal{S} 
^{2}\right\rangle +\frac{1}{2}(\partial _{\mu }G\partial ^{\mu
}G-M_{G}^{2}G^{2})  \nonumber  \label{Leff} \\
&+&c_{d}^{s}\,\left\langle \mathcal{S}\,u_{\mu }\,u^{\mu }\,\right\rangle
+\,c_{m}^{s}\left\langle \mathcal{S}\,\chi _{+}\right\rangle \,\,+\frac{ 
c_{d}^{g}}{\sqrt{3}}G\,\left\langle u_{\mu }\,u^{\mu }\right\rangle \,+\, 
\frac{c_{m}^{g}}{\sqrt{3}}\,G\left\langle \chi _{+}\right\rangle  \nonumber
\\
\hspace*{-0.85cm} &+&c_{e}^{s}\,\left\langle \mathcal{S}\,F_{\mu \nu
}^{+}\,F^{+\,\mu \nu }\right\rangle \,+\frac{c_{e}^{g}}{\sqrt{3}} 
\,G\left\langle \,F_{\mu \nu }^{+}\,F^{+\,\mu \nu }\right\rangle \,+\, 
\mathcal{L}_{\mathrm{mix}}^{P}+\,\mathcal{L}_{\mathrm{mix}}^{S}\,.
\label{L_eff}
\end{eqnarray}%
Here the symbol $\left\langle ...\right\rangle $ denotes the trace over
flavor matrices. The constants $c_{d}^{s}$, $c_{m}^{s}$, $c_{d}^{g}$, 
$c_{m}^{g}$, $c_{e}^{s}$ and $c_{e}^{g}$ define the coupling of scalar fields
and of the bare glueball to pseudoscalar mesons and photons, respectively.
We use the standard notation for the basic blocks 
of the ChPT Lagrangian~\cite{ChPT}: $U=u^{2}=\exp (i\mathcal{P}\sqrt{2}/F)$ 
is the chiral field
collecting pseudoscalar fields in the exponential parametrization, $D_{\mu }$
denotes the chiral and gauge-invariant derivative, \hspace*{0.2cm} $u_{\mu
}=iu^{\dagger }D_{\mu }Uu^{\dagger }$ is the chiral field, $\chi _{\pm
}=u^{\dagger }\chi u^{\dagger }\pm u\chi ^{\dagger }u,\hspace*{0.2cm}\chi
=2B(s+ip),\,\,\,s=\mathcal{M}+\ldots \,$ and $F_{\mu \nu
}^{+}\,=\,u^{\dagger }F_{\mu \nu }Qu+uF_{\mu \nu }Qu^{\dagger }\,,$ where 
$F_{\mu \nu }$ is the stress tensor of the electromagnetic field. The charge
and the mass matrix of current quarks are denoted by $Q=e\, \mathrm{diag} 
\{2/3,-1/3,-1/3\}$ and $\mathcal{M}=\mathrm{diag}\{\hat{m},\hat{m},m_{s}\}$
(we restrict to the isospin symmetry limit with $m_{u}=m_{d}=\hat{m}$); $B$
is the quark vacuum condensate parameter and $F$ the pion decay constant.

The masses of the octet pseudoscalar mesons in the leading order of the
chiral expansion (first term of the Lagrangian) are given by 
\begin{equation}
M_{\pi }^{2}=2\hat{m}B\,,\,M_{K}^{2}=(\hat{m}+m_{s})B\,,
\,M_{\eta ^{8}}^{2}=\frac{2}{3}\,(\hat{m}+2m_{s})B\,.
\end{equation}%
The contribution to the mass of $\eta ^{0}$ in leading order is 
\begin{equation}
\,M_{\eta ^{0}}^{2}=\frac{2}{3}\,(2\hat{m}+m_{s})B\,,
\end{equation}%
i.e. $\eta ^{0}$ is a Goldstone boson in the large $N_{c}$ and in the chiral
limits.

Following~\cite{Ecker:1988te}, we encode in $\mathcal{L}_{\mathrm{mix}}^{P}$
an extra-contribution to the mass of $\eta ^{0}$ (due to the axial anomaly)
and the $\eta ^{0}$-$\eta ^{8}$ mixing term: 
\begin{equation}
\mathcal{L}_{\mathrm{mix}}^{P}=-\frac{1}{2}\gamma _{P}\left( \eta
^{0}\right) ^{2}-z_{P}\eta ^{0}\eta ^{8} \,,
\end{equation} 
(the parameters $\gamma _{P}$ and $z_{P}$ are in turn related to the
parameters $M_{\eta _{1}}$ and $\widetilde{d}_{m}$ of~\cite{Ecker:1988te}).
The physical diagonal states $\eta $ and $\eta ^{\prime }$ are given by 
\begin{equation}
\eta ^{0}\,=\,\eta ^{\prime }\,\cos \theta _{P}\,-\,\eta \,\sin \theta
_{P}\,,\hspace*{0.25cm}\eta ^{8}\,=\,\eta ^{\prime }\,\sin \theta _{P}\,\
+\,\eta \,\cos \theta _{P}\,, 
\end{equation} 
where $\theta _{P}$ is the pseudoscalar mixing angle. We follow the standard
procedure~\cite{Ecker:1988te,Cirigliano:2003yq,Kawarabayashi:1980dp,Venugopal:1998fq} and
diagonalize the corresponding $\eta ^{0}$-$\eta ^{8}$ mass matrix to obtain
the masses of $\eta $ and $\eta ^{\prime }$. By using $M_{\pi }=139.57$~MeV, 
$M_{K}=493.677$~MeV (the physical charged pion and kaon masses), $M_{\eta
}=547.75$~MeV and $M_{\eta ^{\prime }}=957.78$~MeV the mixing angle is
determined as $\theta _{P}=-9.95^{\circ }$, which corresponds to the
tree-level result (see details in Ref.~\cite{Venugopal:1998fq}).
Correspondingly one finds $M_{\eta ^{0}}=948.10$ $MeV$ 
and $z_{P}=-0.105$ GeV $^{2}.$
Higher order corrections in ChPT cause a doubling of the absolute value of
the pseudoscalar mixing angle~\cite{Venugopal:1998fq}); in our work we
restrict to the tree-level evaluation, we therefore consistently use the
corresponding tree-level result of $\theta _{P}=-9.95^{\circ }$.
In the present approach we do not include the neutral pion when considering
mixing in the pseudoscalar sector, because we work in the isospin limit. 
This mixing is small, and can be safely neglected when studying the decay of
scalar resonances into two pseudoscalars. Similarly, for all pseudoscalar
mesons we use the unified leptonic decay constant $F$, which is identified
with the pion decay constant $F=F_{\pi }=92.4$~MeV. A more accurate analysis
including higher orders should use the individual couplings of the
pseudoscalar mesons (for a detailed discussion see 
Refs.~\cite{Gasser:1984gg}).  

\subsection{Scalar quarkonia - glueball mixing}

In this subsection we discuss the glueball-quarkonia mixing. For this reason
we restrict to the following part of the effective Lagrangian (\ref{L_eff}): 
\begin{equation}
\mathcal{L}^{S}=\frac{1}{2}\left\langle D_{\mu }\mathcal{S}D^{\mu }\mathcal{S%
}-M_{\mathcal{S}}^{2}\mathcal{S}^{2}\right\rangle +\frac{1}{2}(\partial
_{\mu }G\partial ^{\mu }G-M_{G}^{2}G^{2})+\mathcal{L}_{\mathrm{mix}}^{S}
\label{L_eff1}
\end{equation}%
where, besides the scalar quarkonia nonet, also the scalar glueball field $G$
has been introduced. In the large $N_{c}$ limit all the states of the
quarkonia nonet have the same nonet mass $M_{\mathcal{S}}$ and the glueball
is decoupled from the quarkonia sector. Deviations from this limit are
encoded in $\mathcal{L}_{\mathrm{mix}}^{S}$, where the glueball-quarkonia
mixing is introduced and where the degeneracy of the nonet states is lifted.

For what concerns the explicit mass term and the next-to-leading order 
$1/N_{c}$ terms in the quarkonia sector we follow~\cite{Cirigliano:2003yq}.
Including additionally a possible breaking of the Gell-Mann-Okubo (GMO) mass
relation and the glueball-quarkonia mixing under the hypothesis of flavor
blindness, we have: 
\begin{equation}
\mathcal{L}_{\mathrm{mix}}^{S}=e_{m}^{S}\left\langle \mathcal{S}^{2}
\chi_{+}\right\rangle +k_{m}^{S}S_{0}\left\langle \chi _{+}\right\rangle - 
\gamma_{S_{0}}\frac{M_{S_{0}}^{2}}{2}S_{0}^{2}-
\gamma _{S_{8}}\frac{M_{S_{8}}^{2}}
{2}S_{8}^{2}-\sqrt{3}fGS_{0}\,.  \label{lsmassmix}
\end{equation} 
The parameter $e_{m}^{S}$ describes the strength of flavor symmetry breaking
derived from the non-zero values of the current quark masses, the parameters 
$k_{m}^{S}$ and $\gamma _{S_{0}}$ describe the order $1/N_{c}$ terms. The
parameter $\gamma _{S_{8}}$ and the related term in the Lagrangian is not
derived in the $N_{c}$ expansion (it is in fact absent 
in~\cite{Cirigliano:2003yq}), but it describes violations of the GMO mass formula
(indeed a result from higher orders in the chiral expansion). Finally, the
parameter $f$ is the glueball-quarkonia mixing strength. Note that $G$,
being a flavor singlet, couples only to the flavor singlet quarkonium state 
$S_{0}$ in the flavor-blind mixing limit. The glueball and the quarkonia
sector decouple in the large $N_{c}$ limit \cite{Lebed:1998st}: a non-zero
mixing, as described by a non-vanishing parameter $f,$ takes into account a
possible deviation from the large $N_{c}$ limit, as the parameters 
$k_{m}^{S} $ and $\gamma _{S_{0}}$ in the quarkonia sector.

The terms contained in $\mathcal{L}_{\mathrm{mix}}^{S}$ lead to mass shifts
of the nonet masses, while also introducing mixing both among quarkonia
states and in the glueball-quarkonia sector. With the use of 
Eqs.~(\ref{L_eff1}) and~(\ref{lsmassmix}) the explicit expression of 
the Lagrangian $\mathcal{L}^{S}$ reads: 
\begin{eqnarray}
\mathcal{L}^{S} &=&-\frac{1}{2}\vec{a}_{0}(\Box +M_{a_{0}}^{2})
\vec{a}_{0}-\,K_{0}^{\ast \,\dagger }\,(\Box +M_{K_{0}^{\ast }}^{2})
\,K_{0}^{\ast }-\frac{1}{2}G(\Box +M_{G}^{2})G  \nonumber \\
&-&\frac{1}{2}M_{S_{8}}^{2}S_{8}^{2}-\frac{1}{2} 
M_{S_{0}}^{2}S_{0}^{2}-z_{S}S_{0}S_{8}-\sqrt{3}fGS_{0}
\label{lsmassexplicit}
\end{eqnarray}%
where $\vec{a}_{0}$ is the isovector collecting $a_{0}^{\pm }$ and 
$a_{0}^{0} $ fields; $K_{0}^{\ast }$ and $K_{0}^{\ast \,\dagger }$ are the
doublets of $(K_{0}^{\ast +}$, $K_{0}^{\ast 0})$ and $(K_{0}^{\ast -}$, 
$\bar{K}_{0}^{\ast 0})$ mesons, respectively (see \ref{appendixA}), and where
the corresponding masses are given by (see Eqs.~(\ref{L_eff1}) 
and~(\ref{lsmassmix})): 
\begin{eqnarray}
M_{a_{0}}^{2} &=&M_{\mathcal{S}}^{2}-4e_{m}^{S}M_{\pi }^{2},  \nonumber \\
M_{K_{0}^{\ast }}^{2} &=&M_{\mathcal{S}}^{2}-4e_{m}^{S}M_{K}^{2},  \nonumber
\\
M_{S_{8}}^{2} &=&M_{\mathcal{S}}^{2}(1+\gamma _{S_{8}})
-\frac{4}{3}e_{m}^{S}(4M_{K}^{2}-M_{\pi }^{2}),  \nonumber \\
M_{S_{0}}^{2} &=&M_{\mathcal{S}}^{2}(1+\gamma _{S_{0}})-\frac{4}{3} 
e_{m}^{S}(2M_{K}^{2}+M_{\pi }^{2}),  \nonumber \\
z_{S} &=&\frac{8\sqrt{2}}{3}\left( e_{m}^{S}+\frac{\sqrt{3}}{2} 
k_{m}^{S}\right) (M_{K}^{2}-M_{\pi }^{2}).  \label{massess0s8}
\end{eqnarray} 
By inverting we get: 
\begin{eqnarray}
M_{\mathcal{S}}^{2} &=&M_{a_{0}}^{2}+\frac{M_{\pi
}^{2}(M_{a_{0}}^{2}-M_{K_{0}^{\ast }}^{2})}{(M_{K}^{2}-M_{\pi }^{2})}, 
\nonumber  \label{Leff_S_1} \\
e_{m}^{S} &=&\frac{M_{a_{0}}^{2}-M_{K_{0}^{\ast }}^{2}}{4(M_{K}^{2}
-M_{\pi}^{2})},  \nonumber \\
\gamma _{S_{8}} &=&\frac{M_{S_{8}}^{2}-M_{\mathcal{S}}^{2}+\frac{4}{3} 
e_{m}^{S}(4M_{K}^{2}-M_{\pi }^{2})}{M_{\mathcal{S}}^{2}},  \nonumber \\
\gamma _{S_{0}} &=&\frac{M_{S_{0}}^{2}-M_{\mathcal{S}}^{2}+\frac{4}{3} 
e_{m}^{S}(2M_{K}^{2}+M_{\pi }^{2})}{M_{\mathcal{S}}^{2}},  \nonumber \\
k_{m}^{S} &=&\frac{2}{\sqrt{3}}\left( \frac{3z_{S}}{8\sqrt{2}  
(M_{K}^{2}-M_{\pi }^{2})}-e_{m}^{S}\right) \,.  \label{paramlsmassmix}
\end{eqnarray} 
We can immediately deduce $M_{\mathcal{S}}$ and $e_{m}^{S}$ for the
considered assignment (which differs from~\cite{Cirigliano:2003yq}) by using
the experimental masses $M_{a_{0}}=M_{a_{0}(1450)}=1.474\pm 0.019$ GeV and 
$M_{K_{0}^{\ast }}=M_{K_{0}^{\ast }(1430)}=1.416\pm 0.006$ 
GeV~\cite{Eidelman:2004wy}: 
\begin{equation}
M_{\mathcal{S}}=1.479\,\,\text{GeV}\,,\;\;e_{m}^{S}=0.199.
\end{equation}

Note that $e_{m}^{S}$ is positive, contrary to the other 
nonets~\cite{Cirigliano:2003yq}. The numerical values for the other constants 
$k_{m}^{S}, $ $\gamma _{S_{0}}$ and $\gamma _{S_{8}}$ (which should be small
if the large $N_{c}$ limit and chiral symmetry still applies approximately
for the scalar nonet) are determined by a fit to data.

Finally we discuss the mass relation in the octet sector. From the first
three equations in (\ref{massess0s8}) we have: 
\begin{equation}
3 M_{S_{8}}^{2} = 4 M_{K_0^{\ast }}^{2} - M_{a_0}^{2} + 
3 \gamma_{S_{8}} M_{\mathcal{S}}^{2},
\end{equation}
where the term proportional to $\gamma _{S_{8}}$ includes deviations form
the GMO limit.

Glueball-quarkonia mixing is usually set up in the basis of fields $N$ 
$\equiv \sqrt{1/2}(\overline{u}u+\overline{d}d)\equiv \overline{n}n$ and $S$ 
$\equiv \overline{s}s$ instead of the octet and singlet fields $S_{0}$ and 
$S_{8}.$ The connection is given by 
\begin{equation}  \label{S_08_to_NS}
S_{0}\,=\,\sqrt{2/3}\,N\,+\,\sqrt{1/3}\,S\,,\hspace*{0.25cm}S_{8}\,=\,
\sqrt{1/3}\,N-\,\sqrt{2/3}\,S\,.  \label{s0s8}
\end{equation}%
Therefore, the scalar-isoscalar part of the Lagrangian~(\ref{lsmassexplicit}) 
involving the fields $N,$ $S$ and $G$ is rewritten as: 
\begin{equation}
\tilde{\mathcal{L}}^{S}\,=\,-\,\frac{1}{2}\,\Phi \,(\Box \,
+\,M_{\Phi}^{2})\,\Phi  \label{mixinglag}
\end{equation} 
with 
\begin{equation}
\Phi =\left( 
\begin{array}{c}
N \\ 
G \\ 
S
\end{array} 
\right) \,,\hspace*{0.3cm}\text{ }M_{\Phi }^{2}=\left( 
\begin{array}{lll}
M_{N}^{2} & \sqrt{2}f & \varepsilon \\ 
\sqrt{2}f & M_{G}^{2} & f \\ 
\varepsilon & f & M_{S}^{2} 
\end{array} 
\right) .  \label{omega}
\end{equation} 
The bare masses $M_{N}$, $M_{S}$ and the flavor mixing parameter 
$\varepsilon $ are determined by inserting (\ref{s0s8}) into 
(\ref{lsmassexplicit}) with: 
\begin{eqnarray}
M_{N}^{2} &=&\frac{2}{3}M_{S_{0}}^{2}+\frac{1}{3}M_{S_{8}}^{2}
+\frac{2\sqrt{2}}{3}z_{S}\,,  \nonumber \\
M_{S}^{2} &=&\frac{1}{3}M_{S_{0}}^{2}+\frac{2}{3}M_{S_{8}}^{2}
-\frac{2\sqrt{2}}{3}z_{S}\,,  \nonumber \\
\varepsilon &=&\frac{\sqrt{2}}{3}\left( M_{S_{0}}^{2}-M_{S_{8}}^{2}\right) - 
\frac{1}{3}z_{S}\,.
\end{eqnarray} 
The inverted relations are: 
\begin{eqnarray}
M_{S_{0}}^{2} &=&\frac{2}{3}M_{N}^{2}+\frac{1}{3}M_{S}^{2}
+\frac{2\sqrt{2}}{3}\varepsilon ,  \nonumber \\
M_{S_{0}}^{2} &=&\frac{1}{3}M_{N}^{2}+\frac{2}{3}M_{S}^{2
}-\frac{2\sqrt{2}}{3}\varepsilon ,  \nonumber \\
z_{S} &=&\sqrt{\frac{2}{3}}\left( M_{N}^{2}-M_{S}^{2}\right) 
- \frac{1}{3}\varepsilon .  \label{ms0ms8zs}
\end{eqnarray} 
As we will show in the next section, as a result of the fit we will
determine the values $M_{N},$ $M_{S},$ and $\varepsilon$, from which we can
deduce $M_{S_{0}}\,,$ $M_{S_{8}}$ and $z_{S}$ or, equivalently 
by~(\ref{paramlsmassmix}), the parameters $k_{m}^{S}\,,$ $\gamma _{S_{0}}$ and 
$\gamma _{S_{8}}$ of the Lagrangian~(\ref{L_eff1}).

The parameter $f$ is the quarkonia-glueball mixing strength, analogous to
the parameter $z$ of 
Refs.~\cite{Amsler:1995tu,Lee:1999kv,Close:2001ga,Burakovsky:1998zg,Strohmeier-Presicek:1999yv}. The mixing strength $z$ refers to the quantum mechanical 
case, where the mass matrix is linear in the bare mass terms. 
The connection between $f$ and $z$, discussed in 
Refs.~\cite{Burakovsky:1998zg,Giacosa:2004ug}, leads to
the approximate relation $f\simeq 2zM_{G}\,.$ 

The parameter $\varepsilon $ induces a direct flavor mixing between the
quarkonia states $N$ and $S$. This effect is neglected 
in~\cite{Amsler:1995tu,Burakovsky:1998zg,Close:2001ga,Strohmeier-Presicek:1999yv},
where flavor mixing is considered to be of higher-order. However, a
substantial $N$-$S$ mixing in the scalar sector is the starting point of the
analysis of Refs.~\cite{Minkowski:2002nf,Minkowski:1998mf}. The origin of
quarkonia flavor mixing is, according 
to~\cite{Minkowski:2002nf,Minkowski:1998mf}, connected to instantons as in the
pseudoscalar channel, but with opposite sign (see also~\cite{Klempt:1995ku}). 
Such a phase structure is also found in the NJL model including the
six-point t' Hooft interaction term~\cite{Hatsuda:1994pi,Klevansky:qe}). 
The mixed physical fields are predicted to be 
a higher lying state of flavor structure 
$[N\sqrt{2}-S]/\sqrt{3}$ and a lower one with $[N+S\sqrt{2}]/\sqrt{3}$. Here
we study the case $\varepsilon \neq 0,$ more precisely $\varepsilon >0,$
which leads to the same phase structure as in 
Ref.~\cite{Minkowski:2002nf,Minkowski:1998mf}, but the quantitative 
results and interpretation will differ.

For the glueball-quarkonia mixing we work in the limit of flavor blindness.
The issue of flavor blindness breaking at the mixing level has been
considered in~\cite{Close:2001ga,Lee:1999kv,Giacosa:2004ug}. This effect can
be taken into account when introducing in the mass mixing matrix $M_{\Phi}^{2}$, 
defined in~(\ref{mixinglag}), an additional parameter $r$ as: 
\begin{equation}
M_{\Phi }^{2}=\left( 
\begin{array}{lll}
M_{N}^{2} & \sqrt{2}f\cdot r & \varepsilon \\ 
\sqrt{2}f\cdot r & M_{G}^{2} & f \\ 
\varepsilon & f & M_{S}^{2}%
\end{array}%
\right) .  \label{omegar}
\end{equation}%
For $r=1$ we regain the original expression, $r\neq 1$ takes into account a
possible deviation from this limit. Determination of this parameter on the
Lattice~\cite{Lee:1999kv} results in $r=1.20\pm 0.07$, in the fit 
of~\cite{Close:2001ga} a value of $r=1\pm 0.3$ is obtained. In the microscopic
quark/gluon model of~\cite{Giacosa:2004ug} the value $r\sim 1.1$-$1.2$ is
deduced. All these findings point to possible small deviations from the
flavor-blind mixing configuration. Hence, in the following we will restrict
to the limit $r=1$.

The orthogonal physical states assigned as $f_{1}\equiv f_{0}(1370)\,,$ 
$f_{2}\equiv f_{0}(1500)$ and $f_{3}\equiv f_{0}(1710)$ resulting from 
Eq.~(\ref{mixinglag}) are obtained by diagonalization of $M_{\Phi }^{2}$ 
(Eq.~(\ref{omega})) with the transformation matrix $B$ as 
\begin{equation}  \label{Transf}
BM_{\Phi }^{2}B^{T}=M_{f}^{2}=\left( 
\begin{array}{lll}
M_{f_{1}}^{2} & 0 & 0 \\ 
0 & M_{f_{2}}^{2} & 0 \\ 
0 & 0 & M_{f_{3}}^{2}%
\end{array}%
\right) ,  \label{b}
\end{equation}%
where the eigenvalues of $M_{f}^{2}$ represent the masses of the physical
states $f_{1}\equiv f_{0}(1370)\,,$ $f_{2}\equiv f_{0}(1500)$ and 
$f_{3}~\equiv~f_{0}(1710)\,.$ 
The physical states $\left\vert i\right\rangle $, 
with $i=f_{1},f_{2},f_{3}$, are then given in terms of the bare states as 
\begin{equation}  \label{fi_to_NGS}
\left\vert i\right\rangle =\sum\limits_{j=N,G,S}B_{ij}\,\left\vert
j\right\rangle \,.  \label{istates}
\end{equation}%
In a covariant framework, where mixing of bound states is studied on a
elementary level, it is not possible to define an orthogonal mixing matrix 
$B $ ($B\cdot B^{T}=B^{T}\cdot B=1_{3}$) connecting the physical to the bare
fields~\cite{Giacosa:2004ug,Hatsuda:1994pi,Klevansky:qe}. 
However, as shown in the model of~\cite{Giacosa:2004ug} 
for the glueball-quarkonia system, deviations
from orthogonality of the mixing matrix $B$ are small, therefore justifying
a Klein-Gordon mixing scenario as in the present approach.

\subsection{Strong and electromagnetic decays of scalar states}

The generic expression for the strong decay width of a scalar state $s$
(both quarkonia and gluonium) into two pseudoscalar mesons $p_{1}$ and 
$p_{2}$ is given by: 
\begin{equation} 
\Gamma _{s\rightarrow p_{1}p_{2}}\,=
\,\frac{\lambda^{1/2}(M_{s}^{2},M_{p_{1}}^{2},M_{p_{2}}^{2})}{16\,\pi \,M_{s}^{3}}
\,\gamma_{sp_{1}p_{2}}\,|M_{s\rightarrow p_{1}p_{2}}|^{2}
\end{equation} 
where $\lambda (x,y,z)=x^{2}+y^{2}+z^{2}-2xy-2yz-2xz$ is the K\"{a}llen
triangle function. The factor $\gamma _{sp_{1}p_{2}}=1/2$ or $1$ stands for
identical or different particles in the final state. In the case of a
degenerate isomultiplet an average over the isospin configurations is
understood.

The matrix elements $M_{s\rightarrow p_{1}p_{2}}$ are expressed in terms of
parameters $c_{d}^{s}$, $c_{m}^{s}$, $c_{d}^{g}$ and $c_{m}^{g}$ (see 
Eq.~(\ref{L_eff})). The parameters $c_{d}^{s}$ and $c_{m}^{s}$ refer to the
scalar quarkonia decays, $c_{d}^{g}$ and $c_{m}^{g}$ to the direct glueball
decays. The complete expressions for the two-pseudoscalar decay widths
(matrix elements included) of the scalar resonances are given in the
Appendix~\ref{spp_widths}.

The decay of the bare glueball states embedded in the physical $f_{0}$
states can proceed in two ways (see Introduction). Mixing expressed by the
parameter $f$ corresponds to the conversion of a bare glueball to a
quarkonia state, which in turn decays into a pseudoscalar meson pair. For 
$f\neq 0$ no physical $f_{0}$ state is a pure glueball, and the decays of 
the quarkonia components are driven by the amplitudes $B_{iN}$ and $B_{iS}$,
which depend on $f$.

The direct decay of the glueball component, without proceeding via an
intermediate scalar quarkonium state, is contained in the parameters 
$c_{d}^{g}$ and $c_{m}^{g}$. For a microscopic description of this mechanisms
we refer to Refs.~\cite{Amsler:1995tu,Strohmeier-Presicek:1999yv}, where the
possible transition of the glueball to two pseudoscalar mesons is described
by processes containing four internal quark/antiquark lines. The parameters 
$c_{d}^{g}$ and $c_{m}^{g}$ are attached to the gluonic amount $B_{iG}$ of
the state $i,$ where $i=f_{1},f_{2},f_{3}$. With the normalization adopted
in Eq. (\ref{L_eff}) the limit $c_{d}^{s}$ $=c_{d}^{g},$ $%
\,c_{m}^{s}=c_{m}^{g}$ refers to a direct glueball decay strength equivalent
to the decay of a flavor-singlet quarkonia state. In the large $N_{c}$ limit
the quarkonia decay constants $c_{d}^{s}$ and $c_{m}^{s}$ scale as $%
N_{c}^{1/2}$, while the glueball decay constants $c_{d}^{g}$ and $c_{m}^{g}$
scale as $N_{c}$~\cite{Lebed:1998st}. Large $N_{c}$ arguments therefore
suggest that the direct glueball decay is suppressed with respect to
quarkonia decays. When performing in the following a fit to data, we first
study the large $N_{c}$ limit by setting the direct glueball decay
parameters $c_{d}^{g}$ and $c_{m}^{g}$ to zero.

The matrix element for the two-photon decay of the scalar $(0^{++})$ state
has the manifestly gauge-invariant form 
\begin{eqnarray}
M_{s \to \gamma \gamma} &=& e^2 \, g_{s\gamma\gamma} \,\, ( g^{\mu\nu} \,
q_1q_2 \, - \, q_1^\nu \, q_2^\mu ) \,\, \epsilon_\mu(q_1) \,
\epsilon_\nu(q_2) \,  \label{Matr_sGG}
\end{eqnarray}
where $q_1$ and $q_2$ are the photon four-momenta, $p = q_1 + q_2$ is the
scalar state momentum and $g_{s\gamma\gamma}$ is the $s\gamma\gamma$
coupling constant. The decay width of the transition $s \to \gamma\gamma$ is
given by 
\begin{eqnarray}  \label{dec_sgg}
\Gamma_{s \to \gamma\gamma} \, = \, \frac{1}{32 \, \pi \, M_s} \,
\sum\limits_{pol} |M_{s \to \gamma \gamma}|^2 \, = \, \frac{\pi}{4} \,
\alpha^2 \, g_{s \gamma\gamma}^2 \, M_s^3\,,
\end{eqnarray}
where $\alpha = e^2/(4\pi) = 1/137$ is the fine structure constant.

The coupling constants $g_{s\gamma \gamma }$ for the bare states 
$S_{0},S_{8},G$ and for the isovector state $a_{0}^{0}$ are directly
calculated using the effective Lagrangian~(\ref{L_eff}): 
\begin{eqnarray}
g_{S_{0}\gamma \gamma }\, &=&\frac{32}{3\sqrt{3}}\,c_{e}^{s}\,,\text{ } 
g_{S_{8}\gamma \gamma }\,=\,\frac{8}{3}\sqrt{\frac{2}{3}}c_{e}^{s}\,\,, 
\hspace*{0.5cm} \\
\text{ }g_{G\gamma \gamma } &=&\frac{32}{3\sqrt{3}}\,c_{e}^{g}\,,\text{ } 
g_{a_{0}\gamma \gamma }\,=\,\frac{8}{3}\sqrt{2}c_{e}^{s} \,. \nonumber
\end{eqnarray} 
The parameter $c_{e}^{s}$ refers to the quarkonia components, while 
$c_{e}^{g}$ contains the direct coupling of the glueball component to the
electromagnetic fields. Latter coupling constant $c_{e}^{g}$ is supposed to
be suppressed, since gluons do not couple directly to the photon field.
However, an intermediate state of two vector mesons for example can in the
framework of vector meson dominance generate a coupling of the glueball to
the two-photon final state; this coupling is supposed to be suppressed and
will not be considered in the numerical analysis. Using the identities for
the field transformations~(\ref{S_08_to_NS}) and~(\ref{fi_to_NGS}) we can
derive the couplings for the bare $N\equiv \sqrt{1/2}(\bar{u}u+\bar{d}d)$
and $S\equiv \bar{s}s$ states and finally for the three physical
scalar-isoscalar states $i=f_{1},f_{2},$ and $f_{3}$: 
\begin{eqnarray} 
g_{N\gamma \gamma }=\frac{5}{\sqrt{2}}\,g_{S\gamma \gamma }=\frac{40}{9}\, 
\sqrt{2}\,\,c_{e}^{s}\,, \ \ \ g_{i\gamma \gamma} = 
\sum\limits_{j=N,G,S} \, B_{ij} \, g_{j\gamma \gamma }\,.
\end{eqnarray} 

\section{Phenomenological fit without direct glueball decay}

\label{fit}

\subsection{General considerations}

In the following we determine a best fit of the parameters entering in 
Eqs.~(\ref{L_eff}) to the experimental averages of masses and decay modes listed
in Ref.~\cite{Eidelman:2004wy}. We first analyze the case of a non-decaying
glueball, i.e. $c_{d}^{g}\,=c_{m}^{g}=0$, where the decays are dominated by
the quarkonia components (as in the original work of \cite{Amsler:1995tu})
in line with large $N_{c}$ arguments. The phenomenological analysis of 
Ref.~\cite{Strohmeier-Presicek:1999yv} confirmed this trend, but, as already
mentioned, the fit of~\cite{Close:2001ga} shows a strong contribution from
the direct decays of the glueball configuration.

The parameters of the model entering in the fit are the three bare masses 
$M_{N},$ $M_{G},$ $M_{S}$, the two mixing parameters $f$ and $\varepsilon $
and the two quarkonia decay parameters $c_{d}^{s}$ and $c_{m}^{s}$: 
\begin{equation}
M_{N},\text{ }M_{G},\text{ }M_{S},\text{ }f,\text{ }\varepsilon ,\text{ } 
c_{d}^{s},\text{ }c_{m}^{s}\, .  \label{param}
\end{equation}

As an experimental input we use the following accepted values 
from~\cite{Eidelman:2004wy}:

(a) The scalar-isoscalar $f_0$ masses with the corresponding values of 
\begin{eqnarray}
M_{f_{1}\equiv f_{0}(1370)} &=&1.35\pm 0.15\text{ GeV}\,,  \label{fit1param}
\\
M_{f_{2}\equiv f_{0}(1500)} &=&1.507\pm 0.005\text{ GeV}\,, \\
M_{f_{3}\equiv f_{0}(1710)} &=&1.714\pm 0.005\text{ GeV}\,.
\end{eqnarray}

(b) The partial decay widths of $f_{2}\equiv f_{0}(1500)$ with:

\begin{eqnarray}
\Gamma _{f_{2}\rightarrow \pi \pi } &=&0.0380\pm 0.0050\text{ GeV}\,, \\
\Gamma _{f_{2}\rightarrow \overline{K}K} &=&0.0094\pm 0.0017\text{ GeV}\,, \\
\Gamma _{f_{2}\rightarrow \eta \eta } &=&0.0056\pm 0.0014\text{ GeV}\,.
\end{eqnarray}

(c) The two accepted ratios for $f_{3}\equiv f_{0}(1710)$: 
\begin{eqnarray}
\Gamma _{f_{3}\rightarrow \pi \pi }/\Gamma _{f_{3}\rightarrow \overline{K}K}
&=&0.20\pm 0.06, \\
\Gamma _{f_{3}\rightarrow \eta \eta }/\Gamma _{f_{3}\rightarrow 
\overline{K} K} &=&0.48\pm 0.15.
\end{eqnarray}

(d) The state $f_{0}(1710)$ has only been observed in the decays into two
pseudo\-scalar mesons~\cite{Eidelman:2004wy}. The decay into the final state 
$4\pi $, which can be fed by higher meson resonances, is 
suppressed~\cite{Barberis:2000cd}. We therefore impose the additional condition 
that the sum of partial decay widths into two pseudoscalar mesons $\left(
\Gamma_{f_{3}}\right)_{2P}$ saturates the total width 
$\left(\Gamma_{f_{3}} \right)_{tot}$: 
\begin{equation}
\left( \Gamma_{f_{3}}\right)_{2P}=\left( \Gamma_{f_{3}}\right)_{tot} =
140\pm 10\text{ MeV}\,.  \label{fit9param}
\end{equation}
Such a constraint is necessary to obtain meaningful total decay widths:
without this condition on the full width a minimum for $\chi^{2}$ is
obtained where $(\Gamma_{f_{3}})_{tot}$ is larger than $1$ GeV, a clearly
unacceptable solution.

(e) The two decay ratios of the $I=1$ state $a_{0}(1450)$: 
\begin{eqnarray}
\frac{\Gamma _{a_{0}\rightarrow \pi \eta ^{\prime }}} {\Gamma_{a_{0} 
\rightarrow \pi \eta }} &=&0.35\pm 0.16,  \label{fit10param} \\
\frac{\Gamma _{a_{0}\rightarrow KK}}{\Gamma_{a_{0}\rightarrow \pi \eta }}
&=&0.88\pm 0.23.  \label{fit11param}
\end{eqnarray}
The total width of $a_{0}(1450)$ into two pseudoscalars is not known,
because of the uncertainty of other decay widths as for $\omega \pi \pi.$

(f) The $I=1/2$ state $K_0^{\ast}(1450)$ decays dominantly into $K\pi $ with
the corresponding width 
\begin{equation}
\Gamma _{K_0^{\ast}\rightarrow K\pi }=273\pm 51\text{ MeV} \,.
\label{fit12param}
\end{equation}

The only accepted average not included in the fit is 
$\Gamma_{f_{2}\rightarrow \eta \eta ^{\prime }}$ for the reason that the 
decay channel $\eta \eta ^{\prime }$ is produced at threshold. Therefore, 
a significant distortion due to the finite width of the state is expected.

For the $N=12$ experimental values listed in Eqs.~(\ref{fit1param})-
(\ref{fit12param}) we perform a $\chi ^{2}$ fit with 
\begin{equation}
\chi ^{2}=\chi ^{2}[M_{N},M_{G},M_{S},f,\varepsilon
,c_{d}^{s},c_{m}^{s}]=\sum_{i=1}^{N=12} \left[ \frac{A_{i}^{theory} 
- A_{i}^{\exp }}{\triangle A_{i}}\right]^2\,,
\end{equation}
where $A_{i}^{\exp}$ represents the i-th experimental result, $\triangle
A_{i}$ is its error and $A_{i}^{theory}$ is the corresponding theoretical
expression, depending on the parameters of~(\ref{param}).

For the case, where the direct glueball is suppressed (i.e. we set 
$c_{d}^{g}=c_{m}^{g}=0$), two local minima for $\chi ^{2}$ are obtained, the
consequences of which we will describe in the following. The first solution
was already analyzed in~\cite{Giacosa:2005qr}, while the second one is a
novel solution with some peculiar characteristics.

\subsection{First solution and implications}

\textit{Fit results:} From the first solution (I) we extract following fit
parameters: 
\begin{eqnarray}
M_{N} &=&1.455~\mathrm{GeV},~M_{G}=1.490~\mathrm{GeV}, 
~M_{S}=1.697~\mathrm{GeV}\,,  \nonumber \\
f&=&0.065~\mathrm{GeV}^{2},~\varepsilon =0.211~\mathrm{GeV}^{2},  \nonumber
\\
~c_{d}^{s}&=&8.48~ \mathrm{MeV},~c_{m}^{s}=2.59~\mathrm{MeV};
~\chi_{tot}^{2}=29.01\,.  \label{fitparam1}
\end{eqnarray}

The corresponding fit results are reported in Table 1.

\begin{center}
\textbf{Table 1.} Fitted mass and decay properties of scalar mesons.

\vspace*{0.5cm} 
\begin{tabular}{|l|l|l|l|}
\hline
Quantity & Exp & Theory & $\chi _{i}^{2}$ \\ \hline
$M_{f_{1}}$ \thinspace (\textrm{MeV}) & $1350$ $\pm 150$ & $1417$ & $0.202$
\\ 
$M_{f_{2}}$ \thinspace (\textrm{MeV}) & $1507\pm 5$ & $1507$ & $\sim 0$ \\ 
$M_{f_{3}}$ \thinspace (\textrm{MeV}) & $1714\pm 5$ & $1714$ & $0.003$ \\ 
$\Gamma _{f_{2}\rightarrow \pi \pi }$ \thinspace \thinspace (\textrm{MeV}) & 
$38.0\pm 4.6$ & $38.52$ & $0.011$ \\ 
$\Gamma _{f_{2}\rightarrow \overline{K}K}$ 
\thinspace \thinspace (\textrm{MeV}) & 
$9.4\pm 1.7$ & $10.36$ & $0.322$ \\ 
$\Gamma _{f_{2}\rightarrow \eta \eta }$ \thinspace (\textrm{MeV}) & $5.6\pm
1.3$ & $1.90$ & $8.109$ \\ 
$\Gamma _{f_{3}\rightarrow \pi \pi }/\Gamma_{f_{3}\rightarrow\overline{K}K}$ 
& $0.20\pm 0.06$ & $0.212$ & $0.036$ \\ 
$\Gamma _{f_{3}\rightarrow \eta \eta }/\Gamma_{f_{3}\rightarrow\overline{K}K}$ 
& $0.48\pm 0.15$ & $0.249$ & $2.446$ \\ 
$\Gamma_{a_{0}\rightarrow \overline{K}K}/
\Gamma_{a_{0}\rightarrow \pi \eta }$ & 
$0.88\pm 0.23$ & $0.838$ & $0.032$ \\ 
$\Gamma_{a_{0}\rightarrow \pi \eta ^{\prime }}/\Gamma _{a_{0}\rightarrow
\pi \eta }$ & $0.35\pm 0.16$ & $0.288$ & $0.150$ \\ 
$\Gamma_{K_{0}^{\ast }\rightarrow K\pi }$ \thinspace (\textrm{MeV}) & 
$273\pm 51$ & $59.10$ & $17.590$ \\ 
$\left(\Gamma _{f_{3}}\right) _{2P}$ \thinspace (\textrm{MeV}) & $140\pm 10$
& $143.27$ & $0.110$ \\ 
$\chi _{tot}^{2}$ & - & - & $29.01$ \\ \hline
\end{tabular}
\end{center}

\textit{Bare masses:} The bare non-strange quarkonia field $N$ has a mass of 
$M_{N}=1.455$~\textrm{GeV}, which is, as desired, similar to the scale set
by the isotriplet combination $a_{0}(1450)$ with a mass of 
$M_{a_{0}}=1.474\pm 0.019$~\textrm{GeV}~\cite{Eidelman:2004wy}. The mass of
the bare glueball $M_{G}=1.490$~\textrm{GeV} is in agreement with the
lattice results~\cite{Michael:2003ai} and with the phenomenological analyses
of~\cite{Amsler:1995tu,Close:2001ga,Strohmeier-Presicek:1999yv}. The bare
state $S$ has a mass of $M_{S}=1.697$~\textrm{GeV}, which is about $\sim 200$
MeV heavier than the $N$ state, an acceptable mass difference like in the
tensor meson nonet.

\textit{Mixing parameters} For the glueball-quarkonia mixing parameter we
get $f=0.065$ GeV$^{2}$, which by the approximate relation 
$f\simeq ~2zM_{G}$~\cite{Burakovsky:1998zg,Giacosa:2004ug} corresponds to 
$z\simeq 21.8$ MeV.

The results of Refs.~\cite{Close:2001ga,Strohmeier-Presicek:1999yv,Giacosa:2004ug} 
are $z=85\pm 10$ MeV, $z=80$ MeV and $z\simeq 62$ MeV, respectively, 
i.e. of the same order, but larger. 
The introduction of additional flavor mixing between the
quarkonia configurations in the fit, as done here, leads to a reduction of
the strength parameter $f$. The lattice result of Ref.~\cite{Lee:1999kv}
with $43\pm 31$ MeV is in agreement with the present evaluation, but has a
large uncertainty. A mixing strength of the same order is found in the
lattice evaluation of~\cite{McNeile:2000xx}.

The flavor mixing parameter resulting from the fit is $\varepsilon =0.211$
GeV$^{2}$. In the limit $f=0$ the mixed physical states are $\left\vert
f_{1}\right\rangle =0.97\left\vert N\right\rangle +0.26\left\vert
S\right\rangle $ and $\left\vert f_{3}\right\rangle =-0.26\left\vert
N\right\rangle +0.97\left\vert S\right\rangle $ (and, of course, $\left\vert
f_{2}\right\rangle =\left\vert G\right\rangle $). The phase structure of the
mixed states is, as discussed previously, as 
in~\cite{Minkowski:2002nf,Minkowski:1998mf,Klempt:1995ku}. 
But here the strength of flavor mixing is smaller, resulting in mixed states, 
which are dominantly $N$ or $S$. The influence however of (an even small) 
flavor mixing in strong and electromagnetic decays may be non-negligible.

\textit{Mixing matrix:} The mixing matrix $B$ relating the physical to the
bare states in the present fit is expressed as: 
\begin{equation}
\,\left( 
\begin{array}{l}
\left\vert f_{1}\right\rangle \equiv \left\vert f_{0}(1370)\right\rangle \\ 
\left\vert f_{2}\right\rangle \equiv \left\vert f_{0}(1500)\right\rangle \\ 
\left\vert f_{3}\right\rangle \equiv \left\vert f_{0}(1710)\right\rangle 
\end{array}
\right) =\left( 
\begin{array}{lll}
0.86 & 0.45 & 0.24 \\ 
-0.45 & 0.89 & -0.06 \\ 
-0.24 & -0.06 & 0.97 
\end{array}
\right) \left( 
\begin{array}{l}
\left\vert N\right\rangle \equiv \left\vert \bar{n}n\right\rangle \\ 
\left\vert G\right\rangle \equiv \left\vert gg\right\rangle \\ 
\left\vert S\right\rangle \equiv \left\vert \bar{s}s\right\rangle 
\end{array}
\right) .  \label{mixingmat1}
\end{equation}

The physical resonances are dominated by the diagonal bare components,
qualitatively in line with 
Refs.~\cite{Amsler:2004ps,Amsler:1995tu,Close:2001ga,Strohmeier-Presicek:1999yv}. 
Since the glueball does not contribute to the decay, the relative phase with
respect to the quarkonia components is at this stage irrelevant. By
inverting $f\rightarrow -f$ we would find the same results for the decays,
but opposite glueball-quarkonia phases. In turn, the relative phases of the 
$N$ and $S$ components are not symmetric under $\varepsilon 
\rightarrow - \varepsilon $. As discussed above, in $f_{0}(1370)$ they are in phase,
while in $f_{0}(1710)$ they are out of phase. The state $\left\vert
f_{0}(1500)\right\rangle $ behaves like a $N$ state with a decreased width,
while the $S$ component is 
small~\cite{Strohmeier-Presicek:1999yv,Amsler:2002ey}. 
Thus the decay into $\overline{K}K$ is smaller than for the $\pi \pi $ channel. 
In the present solution (I) the $N$ and $S$ state components are in phase contrary 
to the results of~\cite{Amsler:1995tu,Strohmeier-Presicek:1999yv}. 
However, the other solution (II), presented later on, shows again an opposite 
phase in $f_{2}=f_{0}(1500),$ but with a large $\overline{s}s$ amount.

\textit{Large $N_{c}$ constants:} From the fit parameters of~(\ref{fitparam1}) 
we determine $\gamma_{S_{0}},$ $\gamma _{S_{8}}$ and $k_{m}^{S}$ by
using~(\ref{ms0ms8zs}) and~(\ref{paramlsmassmix}): 
\begin{eqnarray}
M_{\mathcal{S}} &=&1.479\text{ GeV, }e_{m}^{S}=0.199,  \nonumber \\
\gamma _{S_{8}} &=&0.225,\text{ }\gamma _{S_{0}}=0.236,
\text{ }k_{m}^{S}=-0.818\text{ }\,.
\end{eqnarray}%
The values of $\gamma _{S_{0}}$ and $k_{m}^{S}$ are smaller than 
in~\cite{Cirigliano:2003yq}. Also, the violation of the GMO relation, encoded in 
$\gamma _{S_{8}},$ is small, indicating that higher order corrections in the
chiral expansion are possibly not too large to invalidate the present study.
In the present scenario the smallness of the glueball mixing parameter $f$
can also be interpreted as a small violation of the large $N_{c}$ limit. The
present results show that large $N_{c}$ and chiral symmetry can, although
violated at some level, be a useful guideline to scalar meson physics.

\textit{Resonance $f_{0}(1370)$:} The experimental uncertainties of the 
$f_{0}(1370)$ resonance are large, no average or fit is presented 
in~\cite{Eidelman:2004wy}. The main problem connected with this resonance 
is its large width $(200$-$500$ MeV) and its partial overlap with the broad
low-lying $\sigma \equiv f_{0}(400-1200).$ However, the results from 
WA102~\cite{Barberis:2000cd} indicate a large $N\equiv \overline{n}n$ 
component  in its wave function. Results from Crystal Barrel (summarized 
in~\cite{Amsler:1997up} and subsequently analyzed in~\cite{Abele:2001pv}) 
confirm such a trend (see also~\cite{Thoma:2003in} for a recent review).

\begin{center}
\textbf{Table 2.} Decays of $f_{1}=f_{0}(1370)$.

\vspace*{.5cm} 
\begin{tabular}{|l|l|l|}
\hline
Quantity & Exp (WA102) & Theory \\ \hline
$\Gamma _{f_{1}\rightarrow \overline{K}K}/
\Gamma _{f_{1}\rightarrow \pi \pi }$ & 
$0.46\pm 0.19$ & $0.34$ \\ 
$\Gamma _{f_{1}\rightarrow \eta \eta }/\Gamma _{f_{1}\rightarrow \pi \pi }$
& $0.16\pm 0.07$ & $0.06$ \\ 
$\left( \Gamma _{f_{1}}\right) _{2P}$ (\textrm{MeV}) & \textquotedblright
small\textquotedblright & $166$ \\[1mm] \hline
\end{tabular}
\end{center}

Predictions for the two-pseudoscalar decay modes are in acceptable agreement
with the results of WA102 as shown in Table 2. The measured ratio 
$\Gamma_{f_{1}\rightarrow 4\pi }/ \Gamma _{f_{1}\rightarrow \pi \pi}
=34.0_{-9}^{+22}$~\cite{Barberis:2000cd}, although the errors are large,
points to a dominant $4\pi $ contribution to the total width. Our prediction
gives however a sizable contribution of the two-pseudoscalar decay mode and
is therefore not in agreement with such a large $4\pi $ decay mode.

In the original work of~\cite{Amsler:1995tu}, a quarkonium $\overline{n}n$
state has a very large two-pseudoscalar width with 
$\Gamma_{\overline{n}n\rightarrow \pi \pi }=270\pm 25$~MeV, 
$\Gamma_{\overline{n}n\rightarrow\overline{K}K}=195\pm 20$~MeV 
and $\Gamma_{\overline{n}n\rightarrow \eta\eta}=95\pm 10$~MeV,  
i.e. $(\Gamma_{\overline{n}n})_{2P}\sim 500$~MeV. 
The $f_{1}\equiv f_{0}(1370)$ is in~\cite{Amsler:1995tu} dominantly 
$\overline{n}n,$ therefore one expects a large value for $\left( \Gamma
_{f_{1}}\right) _{2P},$ in contrast to the presented experimental analyses
listed above. A large value for $(\Gamma _{\overline{n}n})_{2P}$ is also
predicted in~\cite{Strohmeier-Presicek:1999yv}: for the mixed state 
$f_{0}(1370)$ one has $\left( \Gamma_{f_{1}}\right) _{2P}=115.7$~MeV
comparable to the present study.

On the contrary, in the study of~\cite{Close:2001ga}, small two-pseudoscalar
partial widths are obtained by the following mechanism: the glueball decay
amplitude in $f_{0}(1370)$ to two pseudoscalar is large and interferes
destructively with the $\overline{n}n$ component (the phases
quarkonia-glueball in~\cite{Close:2001ga} are inverted with respect 
to~(\ref{mixingmat1})). As a result~\cite{Close:2001ga} the two-pseudoscalar decay
width $(\Gamma _{f_{1}})_{2P}$ is smaller than $(\Gamma_{f_{2}})_{2P}$ with 
$(\Gamma_{f_{2}})_{2ps}/(\Gamma _{f_{1}})_{2ps}=10.0\pm 3.0$. At the same
time $(\Gamma_{f_{3}})_{2P}/(\Gamma _{f_{1}})_{2P}=0.7\pm 0.2$ is obtained.

In the present fit with an inert glueball (as in the original work 
of~\cite{Amsler:1995tu} and as in \cite{Strohmeier-Presicek:1999yv}, where 
the glueball is allowed to decay, but the quarkonia components still dominate)
we find the following decay widths into two pseudoscalar pairs: 
\begin{equation} 
(\Gamma _{f_{1}})_{2P}=166\text{ MeV}>(\Gamma _{f_{3}})_{2P}=143\text{MeV} 
>(\Gamma _{f_{2}})_{2P}=51\text{ MeV}. 
\end{equation} 
An analysis by Crystal Barrel~\cite{Abele:2001pv} also indicates sizable
partial decay widths of the $4\pi $ decay channels: 
$\Gamma_{f_{1}\rightarrow \sigma \sigma }=120.5\pm 65$ MeV and 
$\Gamma_{f_{1}\rightarrow \rho \rho }=62.2\pm 28.8$ MeV. 
The same analysis gives the following two-pseudoscalar partial 
widths~\cite{Abele:2001pv,Thoma:2003in}: 
$\Gamma _{f_{1}\rightarrow \pi \pi }=21.7\pm
9.9 $ \textrm{MeV}\thinspace , $\Gamma _{f_{1}\rightarrow 
\overline{K}K}=(7.9\pm
2.7$ \textrm{MeV}) to ($21.2\pm 7.2$ MeV), $\Gamma _{f_{1}\rightarrow \eta
\eta }=0.41\pm 0.27$~MeV. On the contrary, the analysis of~\cite{Bugg:1996ki} 
reports the ratio $\Gamma _{f_{1}\rightarrow \pi \pi }/(\Gamma
_{f_{1}})_{tot}=0.26\pm 0.09,$ pointing to a large $\pi \pi $ (ergo to a
large two-pseudoscalar) partial decay width for $f_{0}(1370).$ This
experimental result is therefore in disagreement with the analysis 
of~\cite{Abele:2001pv}. New results on $f_{0}(1370)$ would be crucial to 
disentangle the scalar puzzle and to understand if a destructive 
glueball/quarkonia interference as in~\cite{Close:2001ga} is necessary.

\textit{Resonance $f_{0}(1500)$:} The theoretical partial widths of the 
$f_{0}(1500)$ are in good agreement with the data (see Table~1) apart from a
slight underestimate of the $2\eta $ channel. We also obtain 
$\Gamma_{f_{2}\rightarrow \eta \eta^{\prime}}=0.036$ MeV as compared to the
experimental value of $\Gamma_{f_{2}\rightarrow \eta \eta ^{\prime }}=2\pm 1$
MeV. Taking into account the finite width of the resonance will lead to an
increase of the theoretical value.

\textit{Resonance $f_{0}(1710)$:} For the decays of $f_{0}(1710)$ we
summarize our results compared to the data of WA102~\cite{Barberis:2000cd}
in Table 3.

\begin{center}
\textbf{Table 3.} Decays of $f_{3}=f_{0}(1710)$.

\vspace*{.5cm} 
\begin{tabular}{|l|l|l|}
\hline
Quantity & Exp (WA102) & Theory \\ \hline
$\Gamma _{f_{3}\rightarrow \bar{K}K}/\Gamma _{f_{3}\rightarrow \pi \pi }$ & 
$5.0\pm 0.7$ & $4.70$ \\ 
$\Gamma _{f_{3}\rightarrow \eta \eta }/\Gamma _{f_{3}\rightarrow \pi \pi }$
& $2.4\pm 0.6$ & $1.17$ \\ 
$\Gamma _{f_{3}\rightarrow \eta \eta ^{\prime }}/\Gamma _{f_{3}\rightarrow
\pi \pi }$ & $<0.18$ & $1.59$ \\ 
$\left( \Gamma _{f_{3}}\right) _{2P}$ (\textrm{MeV}) & \textquotedblright
dominant\textquotedblright & $143.27$ \\[1mm] \hline
\end{tabular}
\end{center}

The first two ratios, already included in the fit of Table 2, can be
reproduced. The theoretical ratio $\Gamma _{f_{3}\rightarrow \eta
\eta^{\prime }}/ \Gamma _{f_{3}\rightarrow \pi \pi }$, which is not included
in~\cite{Eidelman:2004wy}, is in complete disagreement with the WA102
result. The dominance of the $\eta \eta ^{\prime }$ mode over $\pi \pi $ is
a solid prediction in the framework of solution I, which does not depend
very much on the choice of parameters. A confirmation of the experimental
result could possibly hint at a different mixing scenario or at a sizable
role of direct glueball decay.

\textit{Resonance $a_{0}(1450)$:} The ratios of two-pseudoscalar decay modes
of $a_{0}(1450)$, included in the fit of Table 2, are well reproduced. The
prediction for the two-pseudoscalar width of $(\Gamma _{a_{0}})_{2P}=\ 84.26$
MeV is smaller than the total width of $265\pm 13$ MeV. However, the
experimental ratio ($\Gamma_{a_{0}\rightarrow\omega\pi\pi
}/\Gamma_{a_{0}\rightarrow\pi\eta})$ is not known: no average or fit is
listed in~\cite{Eidelman:2004wy}. The experimental value 
from~\cite{Baker:2003jh}, which is $10.7\pm 2.3$, 
would imply a dominant $\omega \pi
\pi $ mode and in turn a rather small two-pseudoscalar partial decay widths.
This finding is in disagreement with the results 
of~\cite{Amsler:1995tu,Gobbi:1993au}. In~\cite{Amsler:1995tu} a value of 
$(\Gamma_{a_{0}})_{2P}~=~390~\pm~110$~MeV is found, 
in the work of~\cite{Gobbi:1993au}
one has $\left( \Gamma _{a_{0}}\right) _{2P} = 420 \, - \, 940$~MeV for $a_{0}$
masses in the range of $1200 \, - \, 1400$~MeV. For our value for $(\Gamma
_{a_{0}})_{2P}$ we obtain the estimate: $\Gamma _{a_{0}\rightarrow \omega
\pi \pi }/ \Gamma _{a_{0}\rightarrow \pi \eta }=[
(\Gamma_{a_{0}})_{tot}-(\Gamma_{a_{0}})_{2P}]/\ \Gamma _{a_{0}\rightarrow
\pi \eta}\sim~4.5\,.$

\textit{Resonance $K_0^{\ast}(1430)$:} Our result for $\Gamma
_{K_0^{\ast}\rightarrow K\pi }$ underestimates the experimental value by a
factor of about 5 (see Table 1). Furthermore, for the additional $K\eta $
decay channel we get $\Gamma_{K_0^{\ast}\rightarrow K\eta }/\Gamma
_{K_0^{\ast} \rightarrow \pi K}=0.026$.

In~\cite{Gobbi:1993au} a value of $\Gamma _{K_{0}^{\ast }\rightarrow K\pi
}=340$ MeV is predicted, but, as discussed above, $\left( \Gamma
_{a_{0}}\right) _{2P}$ is of the order of $1$ GeV, much larger than the full
width. Similarly, in~\cite{Amsler:1995tu} with 
$\Gamma _{K_{0}^{\ast}\rightarrow K\pi }=200\pm 20$ and 
$(\Gamma _{a_{0}})_{2P}=390\pm 110$ MeV
the first result underestimates while the second overshoots the experimental
value. A full analysis in the $^{3}P_{0}$ model~\cite{Ackleh:1996yt} results
in $\Gamma _{K^{\ast }\rightarrow \pi K}=166$~MeV and $\Gamma _{N\rightarrow
\pi \pi }=271$ MeV; unfortunately the resonance $a_{0}(1450)$ is not
discussed in~\cite{Ackleh:1996yt}. The authors of \cite{Ackleh:1996yt} also
tried to adjust $\Gamma _{K_{0}^{\ast }\rightarrow \pi K}$ to its
experimental value, and then calculate the $2\pi $ partial width of a $N$
state, obtaining $\Gamma _{N\rightarrow \pi \pi }\sim 450$ MeV. The last
result implies a very large two-pseudoscalar and full width for a $N$ state.
A full experimental determination of all relevant decay modes involving 
$a_{0}(1450)$ and $K_{0}^{\ast }(1430)$ would certainly help to clarify this
issue. We refer to section \ref{k0} for a further discussion of this problem.

If $f_{0}(1370)$ is dominantly $\bar{n}n,$ as 
in~\cite{Amsler:1995tu,Close:2001ga,Lee:1999kv,Burakovsky:1998zg,Strohmeier-Presicek:1999yv,Giacosa:2004ug}, there is, as discussed above, an incompatibility 
of the present experimental small two-pseudoscalar partial decay 
widths~\cite{Abele:2001pv} and various model calculations. 
At the same time, a consistent understanding of the isodoublet states 
$K_0^{\ast}(1450)$ and the isovectors $a_{0}(1450)$ is still incomplete.

\textit{Two-photon decays:} As a further consequence we discuss the
two-photon decay rates of the scalar resonances. We assume that the coupling 
$c_{e}^{g}$ is suppressed with respect to $c_{e}^{s},$ i.e. we set the
glueball-photon coupling $c_{e}^{g}$ to zero. The ratios of radiative decay
widths as a prediction of the fit are: 
\begin{equation}
\Gamma _{f_{1}\rightarrow 2\gamma }:\Gamma _{f_{2}\rightarrow 2\gamma
}:\Gamma _{f_{3}\rightarrow 2\gamma }:\Gamma _{a_{0}^{0}\rightarrow 2\gamma
}=1:0.305:0.002:0.471\,,  \label{sol1twophot}
\end{equation}%
which are independent of the coupling $c_{e}^{s}$. The result for 
$\Gamma_{f_{2}\rightarrow 2\gamma }/\Gamma _{f_{1}\rightarrow 2\gamma }$ is in
qualitative agreement with the results of~\cite{Close:2001ga,Giacosa:2004ug}. 
The ratio $\Gamma _{f_{3}\rightarrow 2\gamma }/\Gamma _{f_{1}\rightarrow 2\gamma }$, 
however, is considerably smaller than in the previous works. The
suppression of $\Gamma _{f_{3}\rightarrow 2\gamma }$ originates from the
destructive interference between the $N$ and $S$ components, which in turn
is traced to the flavor mixing with $\varepsilon >0$ in accord with the
phases of~\cite{Minkowski:2002nf,Minkowski:1998mf}. Another interesting
prediction is the ratio $\Gamma _{a_{0}^{0}\rightarrow 2\gamma }/\Gamma_{f_{1}
\rightarrow 2\gamma }$, which is relatively large.

The experimental status of the two-photon decays is still incomplete. For
the $f_{0}(1370)$ two values are indicated in PDG2000~\cite{Groom:in}: 
$3.8\pm 1.5$ $\mathrm{keV}$ and $5.4\pm 2.3$ keV. However, it is not clear if
the two-photon signal comes from the $f_{0}(1370)$ or from the high mass end
of the broad $f_{0}(400-1200).$ The PDG currently~\cite{Eidelman:2004wy}
seems to favor this last possibility, but the data could also be valid for
the $f_{0}(1370).$ We therefore interpret the two experimental values as an
upper limit for the two-photon decay width of the $f_{0}(1370).$ Signals for
two-photon decays of $f_{0}(1500)$ and $f_{0}(1710)$ have not yet been seen,
the following upper limits are reported~\cite{Eidelman:2004wy}: 
\begin{eqnarray}
\Gamma _{f_{0}(1500)\rightarrow 2\gamma }(\Gamma _{f_{0}(1500)\rightarrow
\pi \pi }/\Gamma _{f_{0}(1500)tot}) &<&0.46~\mathrm{keV},  \nonumber \\
\Gamma _{f_{0}(1710)\rightarrow 2\gamma }(\Gamma _{f_{0}(1710)\rightarrow K 
\bar{K}}/\Gamma _{f_{0}(1710)tot}) &<&0.11~\mathrm{keV.}
\end{eqnarray}
Using the known branching ratio $\Gamma _{f_{0}(1500)\rightarrow 2\pi}/
\Gamma _{f_{0}(1500)tot}$ one gets $\Gamma _{f_{0}(1500)\rightarrow 2\gamma}<1.4$ 
$\mathrm{keV}$\textrm{\ }~\cite{Amsler:2002ey}. An accepted fit for 
$\Gamma _{f_{0}(1710)\rightarrow K\bar{K}}/\Gamma _{f_{0}(1710)tot}$ is not
reported in \cite{Eidelman:2004wy}. Using the value 
from~\cite{Longacre:1986fh} with $\Gamma _{f_{0}(1710)\rightarrow K\bar{K} }/
\Gamma_{f_{0}(1710)tot}= 0.38_{-0.13}^{+0.03}$ we find an upper limit of
the order of $\Gamma _{f_{0}(1710)\rightarrow 2\gamma }\sim 0.3$ keV.

For an absolute prediction of the two-photon decay widths we use 
$c_{e}^{s}=0.0138$ \textrm{GeV}$^{-1}$ as determined in the model approach 
of Ref.~\cite{Giacosa:2004ug} For the non-strange quarkonium state we get 
$\Gamma _{N\rightarrow 2\gamma }=0.969$ keV, while for the isovector and
mixed scalars we have: 
\begin{eqnarray}
\Gamma _{f_{1}\rightarrow 2\gamma } &=&0.703~\mathrm{keV}\,,\,\,\Gamma
_{f_{2}\rightarrow 2\gamma }=0.235~\mathrm{keV}\,,  \nonumber \\
\Gamma _{f_{3}\rightarrow 2\gamma } &=&0.002~\mathrm{keV}\,,\,\,\Gamma
_{a_{0}^{0}\rightarrow 2\gamma }=0.362~\mathrm{keV}\,.
\end{eqnarray}%
The results for the mixed states are below the current upper limits 
(the original results presented in
\cite{Giacosa:2005qr} contain a slight missprint, the correct 
numbers are reported here.
The physical considerations are not affected 
from this slight change).

The estimate for the $2\gamma $ decay of the bare quarkonium state $N\equiv 
\bar{n}n$ of $0.969$ keV is smaller than the one of~\cite{Amsler:2002ey},
where the following expression has been used: 
\begin{equation}
\Gamma_{\overline{n}n\rightarrow 2\gamma }(0^{++})=k\left( 
\frac{M_{N}(0^{++})}{M_{N}(2^{++})}\right) ^{3} 
\Gamma_{\overline{n}n\rightarrow 2\gamma }(2^{++})
\end{equation} 
The coefficient $k$ is $15/4$ in a non-relativistic calculation, but becomes 
smaller when considering relativistic corrections~\cite{Li:1990sx}. 
In~\cite{Amsler:2002ey} a range of values for $k$ from $2$ to $15/4$ 
is considered.
Our chiral Lagrangian approach combined with~\cite{Giacosa:2004ug} points to
a smaller value of $k$. Using our result for $\Gamma _{N\rightarrow 2\gamma
}(0^{++})$ and taking the value $\Gamma _{N\rightarrow 2\gamma
}(2^{++})=2.60\pm 0.24$ keV~\cite{Eidelman:2004wy} at $M_{N}(2^{++})=1.27$
GeV we get $k\sim 0.25.$ This result is model dependent, since it relies on
the parameters for the covariant description of the scalar mesons used 
in~\cite{Giacosa:2004ug}. A fully covariant treatment may imply strong
deviations from the non-relativistic limit.

\textit{Discussion:} The largest contribution to $\chi^{2}$ is due to the
underestimate of the $K_0^{\ast }\rightarrow K\pi $ width: from Table 1 we
have $\chi_{tot}^{2}/N=2.42.$ When excluding the data point for 
$\Gamma_{K_0^{\ast }\rightarrow K\pi }$ in the fit, a very similar minimum
compared to~(\ref{fitparam1}) is found with: 
\begin{eqnarray}
\hspace*{-0.75cm}M_{N} &=&1.442~\mathrm{GeV}\,,\,\, M_{G}=1.485~\mathrm{GeV}\,, 
\,\, M_{S}=1.695~\mathrm{GeV}\,, \,\, f=0.080~\mathrm{GeV}^{2}\,, 
\nonumber \\
\hspace*{-0.75cm}\varepsilon &=&0.225~\mathrm{GeV}^{2}\,, \,\,
c_{d}^{s}=8.12~\mathrm{MeV}\,, \,\, c_{m}^{s}=3.57~\mathrm{MeV}\,; \,\, \chi
_{tot}^{2}=11.19\,.  \label{fitparam1nok}
\end{eqnarray}
In this case we have $\chi_{tot}^{2}/N=1.02,$ corresponding to a good
description of data. The discussion about the isoscalar states and 
$a_{0}(1450)$ remains unchanged.

The underestimate of $K_{0}^{\ast }\rightarrow K\pi $ constitutes an open
problem of the scalar analysis. As already discussed, a consistent
understanding of the complete scalar nonet is lacking in other approaches as
well. We will further discuss this issue in section~\ref{k0}.

Aside from this difficulty, the rest of the accepted data 
in~\cite{Eidelman:2004wy} is well described. The quality of the current fit
seemingly excludes a sizable direct decay of the scalar glueball component.
Concerning results for data, which are not reported as average or fit 
in~\cite{Eidelman:2004wy}, the situation is less clear: the predicted full
two-pseudoscalar width of $f_{1}=f_{0}(1370)$ is large, when confronted with
the WA102 result (see Table 2 and 
Refs.~\cite{Amsler:1997up,Abele:2001pv,Thoma:2003in}, 
but also the different result of~\cite{Bugg:1996ki}). 
The ratio $\Gamma _{f_{3}\rightarrow \eta \eta ^{\prime}}/
\Gamma _{f_{3}\rightarrow \pi \pi }$ (Table 3) is also problematic in the
present solution. An accepted average, in particular for these values, would
help in clarifying these issues.

\subsection{Second solution and implications}

\textit{Fit results:} Solution II is obtained for following fit parameters: 
\begin{eqnarray}
\hspace*{-0.75cm} M_{N} &=&1.298~\mathrm{GeV}\,, \,\, 
M_{G}=1.513~\mathrm{GeV}\,, \,\, M_{S}=1.593~\mathrm{GeV}\,, \,\, 
f=0.400~\mathrm{GeV}^{2}\,, \nonumber \\
\hspace*{-0.75cm}\varepsilon &=&0.015~\mathrm{GeV}^{2}\,, \,\,
c_{d}^{s}=7.48~\mathrm{MeV}\,, \,\, c_{m}^{s}=6.42~\mathrm{MeV}\,; \,\, 
\chi_{tot}^{2}=24.61\,.  \label{fitparam2}
\end{eqnarray}

The corresponding results are reported in Table 4.

\begin{center}
\textbf{Table 4.} Fitted mass and decay properties of scalar mesons.

\vspace*{0.4cm} 
\begin{tabular}{|l|l|l|l|}
\hline
Quantity & Exp & Theory & $\chi _{i}^{2}$ \\ \hline
$M_{f_{1}}$ \thinspace (\textrm{MeV}) & $1350$ $\pm 150$ & $1142$ & $1.924$
\\ 
$M_{f_{2}}$ \thinspace (\textrm{MeV}) & $1507\pm 5$ & $1508$ & $0.023$ \\ 
$M_{f_{3}}$ \thinspace (\textrm{MeV}) & $1714\pm 5$ & $1713$ & $0.0254$ \\ 
$\Gamma _{f_{2}\rightarrow \pi \pi }$ \thinspace \thinspace (\textrm{MeV}) & 
$38.0\pm 4.6$ & $37.31$ & $0.019$ \\ 
$\Gamma _{f_{2}\rightarrow \overline{K}K}$ 
\thinspace \thinspace (\textrm{MeV}) & 
$9.4\pm 1.7$ & $10.08$ & $0.167$ \\ 
$\Gamma _{f_{2}\rightarrow \eta \eta }$ \thinspace (\textrm{MeV}) & $5.6\pm
1.3$ & $4.70$ & $0.477$ \\ 
$\Gamma _{f_{3}\rightarrow \pi \pi }/\Gamma _{f_{3}\rightarrow\overline{K}K}$ & 
$0.20\pm 0.06$ & $0.216$ & $0.071$ \\ 
$\Gamma _{f_{3}\rightarrow \eta \eta }/\Gamma _{f_{3}\rightarrow\overline{K}K}$
& $0.48\pm 0.15$ & $0.248$ & $2.400$ \\ 
$\Gamma _{a_{0}\rightarrow \overline{K}K}/
\Gamma _{a_{0}\rightarrow \pi \eta }$ & 
$0.88\pm 0.23$ & $1.078$ & $0.741$ \\ 
$\Gamma _{a_{0}\rightarrow \pi \eta ^{\prime }}/\Gamma _{a_{0}\rightarrow
\pi \eta }$ & $0.35\pm 0.16$ & $0.291$ & $0.134$ \\ 
$\Gamma _{K_{0}^{\ast }\rightarrow K\pi }$ \thinspace (\textrm{MeV}) & 
$273\pm 51$ & $53.75$ & $18.480$ \\ 
$\left( \Gamma _{f_{3}}\right) _{2P}$ \thinspace (\textrm{MeV}) & $140\pm 10$
& $143.94$ & $0.155$ \\ 
$\chi _{tot}^{2}$ & - & - & $24.61$ \\ \hline
\end{tabular}
\end{center}
A first look to~(\ref{fitparam2}) shows some peculiar differences when
compared to the set of (\ref{fitparam1}). In the following we discuss the
implications and the differences of this second solution.

\textit{Bare masses:} The bare masses $M_{N}$ and $M_{S}$ are smaller than
in solution I, their mass difference is still around $200$~GeV 
(Eq.~(\ref{fitparam2})). The bare glueball mass is about $\sim $ $1.5$ GeV, 
as before, but now is much closer to $M_{S}$. This small mass difference 
leads to a strong mixing between the glueball and the bare 
$S\equiv \overline{s}s$ state.

\textit{Mixing parameters:} The quarkonia flavor mixing 
$\varepsilon =0.015$ GeV$^{2}$ is very small in this solution 
and has practically no influence on
the phenomenology. On the contrary $f=0.400$ GeV$^{2}$ is much larger,
leading to a strong glueball-quarkonia mixing. In this respect there is a
clear difference between the two solutions. Using the approximate relation 
$f\simeq 2zM_{G}$~\cite{Burakovsky:1998zg,Giacosa:2004ug} we find in this
case $z\simeq 130$ MeV$,$ which is larger than the results from other works
listed in Sec.~III.B.3, but still in qualitative agreement.

\textit{Mixing matrix:} The mixing matrix $B$, relating the physical to the
bare states, for the second solution is expressed as: 
\begin{equation}
\,\left( 
\begin{array}{l}
\left\vert f_{1}\right\rangle \equiv \left\vert f_{0}(1370)\right\rangle \\ 
\left\vert f_{2}\right\rangle \equiv \left\vert f_{0}(1500)\right\rangle \\ 
\left\vert f_{3}\right\rangle \equiv \left\vert f_{0}(1710)\right\rangle
\end{array}
\right) =\left( 
\begin{array}{lll}
0.81 & 0.54 & 0.19 \\ 
-0.49 & 0.49 & 0.72 \\ 
-0.30 & -0.68 & 0.67 
\end{array}
\right) \left( 
\begin{array}{l}
\left\vert N\right\rangle \equiv \left\vert \bar{n}n\right\rangle \\ 
\left\vert G\right\rangle \equiv \left\vert gg\right\rangle \\ 
\left\vert S\right\rangle \equiv \left\vert \bar{s}s\right\rangle 
\end{array}
\right) .
\end{equation}
The large mixing parameter $f$ causes the glueball configuration to be
spread out among the $f_0$ states: $f_{1}\equiv f_{0}(1370)$ is still
dominantly $\bar{n}n$, $f_{2}\equiv f_{0}(1500)$ is mostly $\bar{s}s,$ but
with a sizable out-of-phase $\bar{n}n$ amplitude (the opposite phase of 
$\bar{n}n$ and $\bar{s} s$ was first considered in~\cite{Amsler:1995tu} as a
mechanism to explain the large $\pi \pi /\overline{K}K$ ratio). In 
$f_{3}\equiv f_{0}(1710)$ both the gluonium and the $\bar{s}s$ components are
large, where the gluonic component is slightly larger. 
Remarkably, although the bare level ordering is still $M_{N}<M_{G}<M_{S}$,
the largest $\bar{s}s$ amount is contained in $f_{2}$, while the largest
gluonic component is present in $f_{3}.$
The mixing matrix resembles some features of the 
results of~\cite{Lee:1999kv}, although the bare level ordering is different.

\textit{Large $N_{c}$ constants:} From the parameters in (\ref{fitparam2})
we determine $\gamma _{S_{0}},$ $\gamma _{S_{8}}$ and $k_{m}^{S}$ by 
using~(\ref{paramlsmassmix}) and~(\ref{ms0ms8zs}):  
\begin{eqnarray}
M_{\mathcal{S}} &=&1.479\text{ GeV, }e_{m}^{S}=0.199,  \nonumber \\
\gamma _{S_{8}} &=&0.139,\text{ }\gamma _{S_{0}}= - 0.032,\text{ }
k_{m}^{S}= - 0.786.
\end{eqnarray}
The same considerations as for solution~I also hold here.

\textit{Resonance $f_{0}(1370)$:} The results for $f_{1}=f_{0}(1370)$ are
summarized in Table 5.

\begin{center}
\textbf{Table 5.} Decays of $f_{1}=f_{0}(1370)$.

\vspace*{.5cm} 
\begin{tabular}{|l|l|l|}
\hline
Quantity & Exp (WA102) & Theory \\ \hline
$\Gamma_{f_{1}\rightarrow \overline{K}K}/\Gamma_{f_{1}\rightarrow \pi \pi }$ & 
$~0.46\pm 0.19$ & $0.27$ \\ 
$\Gamma_{f_{1}\rightarrow \eta \eta }/\Gamma_{f_{1}\rightarrow \pi \pi }$ & 
$0.16\pm 0.07$ & $0.02$ \\ 
$\left( \Gamma_{f_{1}}\right)_{2P}$ (\textrm{MeV}) & \textquotedblright
small\textquotedblright & $56.79$ \\[1mm] \hline
\end{tabular}
\end{center}

The main difference with respect to solution I concerning $f_{1}=f_{0}(1370)$
is the decreased theoretical two-pseudoscalar width (mostly caused by the
smaller physical mass, see Table 4). In the present fit we find the
following decay widths into two pseudoscalar pairs: 
\begin{eqnarray}
(\Gamma_{f_{3}})_{2P}=144\text{ MeV}>(\Gamma_{f_{1}})_{2P}=57\text{ MeV}>
(\Gamma _{f_{2}})_{2P}=52\text{ MeV} \,. 
\end{eqnarray} 
Here the decay pattern is in better agreement with the analysis 
of~\cite{Abele:2001pv,Thoma:2003in} than the one of solution I.

\textit{Resonance $f_{0}(1500)$:} The theoretical partial widths of 
$f_{0}(1500)$ are in rather good agreement with the data (see Table 2). We
also obtain $\Gamma _{f_{2}\rightarrow \eta \eta ^{\prime }}=1.5$ MeV as
compared to the experimental value of $\Gamma _{f_{2}\rightarrow \eta
\eta^{\prime }}=2\pm 1$ MeV without invoking further threshold corrections.

\textit{Resonance $f_{0}(1710)$:} For the decays of $f_{0}(1710)$ we
summarize our results compared to the data of WA102~\cite{Barberis:2000cd}
in Table 6.

\begin{center}
\textbf{Table 6.} Decays of $f_{3}=f_{0}(1710)$.

\vspace*{.5cm} 
\begin{tabular}{|l|l|l|}
\hline
Quantity & Exp (WA102) & Theory \\ \hline
$\Gamma _{f_{3}\rightarrow \overline{K}K}/\Gamma _{f_{3}\rightarrow \pi\pi}$ & 
$5.0\pm 0.7$ & $4.63$ \\ 
$\Gamma _{f_{3}\rightarrow \eta \eta }/\Gamma _{f_{3}\rightarrow \pi \pi }$
& $2.4\pm 0.6$ & $1.15$ \\ 
$\Gamma _{f_{3}\rightarrow \eta \eta ^{\prime }}/\Gamma _{f_{3}\rightarrow
\pi \pi }$ & $<0.18$ & $0.36$ \\ 
$\left( \Gamma _{f_{3}}\right)_{2P}$ (\textrm{MeV}) & \textquotedblright
dominant\textquotedblright & $143.94$ \\[2mm] \hline
\end{tabular}
\end{center}

The theoretical $\eta \eta ^{\prime }/\pi \pi $ ratio is now smaller than in
solution~I. Although the prediction is still larger than the upper limit set
by WA102, it does not represent such an evident mismatch.

\textit{Resonance $a_{0}(1450)$:} The two experimental ratios are
satisfactorily described. Otherwise, the discussion of solution~I is still
valid here.

\textit{Resonance $K_0^{*}(1430)$:} The underestimate of $\Gamma_{K_0^{\ast
}\rightarrow K\pi }$ is also present in the second solution (see Table 4).
Furthermore, for the additional $K\eta $ decay channel we get 
$\Gamma_{K_0^{\ast}\rightarrow K\eta }/ \Gamma_{K_0^{\ast}\rightarrow 
\pi K}=0.05065.$

\textit{Two-photon decays:} The two-photon decay ratios resulting from
solution II read: 
\begin{equation}
\Gamma _{f_{1}\rightarrow 2\gamma }:\Gamma _{f_{2}\rightarrow 2\gamma
}:\Gamma _{f_{3}\rightarrow 2\gamma }:\Gamma _{a_{0}^{0}\rightarrow 2\gamma
}=1:0.253:0.055:0.493.
\end{equation}%
The two $\gamma $ results are in qualitative agreement 
with~\cite{Close:2001ga,Giacosa:2004ug}. 

Following the arguments of the previous section by using the result 
of~\cite{Giacosa:2004ug} we find: 
\begin{eqnarray}
\Gamma _{f_{1}\rightarrow 2\gamma } &=&0.350~\mathrm{keV}\,,\,\,\Gamma
_{f_{2}\rightarrow 2\gamma }=0.088~\mathrm{keV}\,,  \nonumber \\
\Gamma _{f_{3}\rightarrow 2\gamma } &=&0.019~\mathrm{keV}\,,\,\,\Gamma
_{a_{0}^{0}\rightarrow 2\gamma }=0.172~\mathrm{keV}\,.
\end{eqnarray}%
As before, the rates are below the upper limits set by current data, but
they depend on the model-dependent evaluation of~\cite{Giacosa:2004ug}.

\textit{Discussion:} If in this case we exclude the $K_{0}^{\ast
}\rightarrow K\pi $ mode from the fit (which generates the by far largest
contribution to $\chi ^{2}$), the following set of parameters is found: 
\begin{eqnarray}
\hspace*{-0.75cm}M_{N} &=&1.313~\mathrm{GeV},
~M_{G}=1.511~\mathrm{GeV},~M_{S}=1.594~\mathrm{GeV},~f=0.395~\mathrm{GeV}^{2}\,,  
\nonumber \\
\hspace*{-0.75cm}\varepsilon &=&0.002~\mathrm{GeV}^{2},~c_{d}^{s}=7.30~
\mathrm{MeV},~c_{m}^{s}=6,61~\mathrm{MeV};~\chi _{tot}^{2}=5.97\,.
\end{eqnarray}%
We then obtain $\chi _{tot}^{2}/N=0.54,$ corresponding also in this case 
to a good description of the remaining data. Again, there is no
phenomenological evidence for a direct decay of the glueball component.

Apart from the decay $K_0^{\ast}\rightarrow K\pi ,$ both solutions (I and
II) describe the data (\ref{fit1param})-(\ref{fit11param}) well. There are,
however, differences when comparing to the WA102 results. The second
solution comes in this respect closer to these data, but the experimental
results are not yet conclusive.

\subsection{$K_{0}^{\ast }(1430)$}

\label{k0}

The most striking mismatch with the data is in both analyzed scenarios I and
II the theoretical underestimate of the $K_{0}^{\ast }\rightarrow K\pi $
mode (Tables 1 and 4). The corresponding partial $\chi _{K_{0}^{\ast
}\rightarrow K\pi }^{2}$ is by far the dominant contribution to $\chi
_{tot}^{2}.$ This mismatch also holds when including direct glueball decays
in the analysis as shown in the next section.

As a further attempt, following~\cite{Ackleh:1996yt}, we can also pursue
another strategy: in a first fit we fix the quarkonium decay parameters 
$c_{d}^{s}$ and $c_{m}^{s}$ in order to reproduce $K_{0}^{\ast }\rightarrow
K\pi $~(Eq. (\ref{fit12param})) and the two ratios for $a_{0}$ given 
in~(\ref{fit10param}) and~(\ref{fit11param}). Then we obtain $c_{d}^{s}=17.94$ 
MeV (larger by a factor 2 when compared to the previous fit in Tables 1 and 4).
and $c_{m}^{s}=7.35$ MeV. With these values one has 
$\Gamma _{K_{0}^{\ast}\rightarrow \pi K}=281$ MeV, 
$\Gamma_{a_{0}\rightarrow \overline{K}K}/
\Gamma_{a_{0}\rightarrow \pi\eta}=0.88$ 
and $\Gamma _{a_{0}\rightarrow \pi \eta^{\prime }}/\Gamma _{a_{0}\rightarrow 
\pi \eta }=0.30$ and $\Gamma_{a_{0}\rightarrow 2P}=381.17$ MeV 
(already larger than the experimental result).

The corresponding decay width of the $N$ state into two pseudoscalars is 
$\Gamma_{N\rightarrow 2P}\sim 900$ MeV (for a mass of $M_{N}~\sim~1.4$~GeV).
We then find similar results as in~\cite{Amsler:1995tu,Ackleh:1996yt},
implying that the discussed trend is rather model independent.
If at this stage we fix $c_{d}^{s}=17.94$ MeV and $c_{m}^{s}=7.35$ MeV and
we make a second fit of the remaining free parameters $(M_{N},$ $M_{G},$ 
$M_{S},$ $f,$ $\varepsilon ,$ $c_{d}^{g},$ $c_{m}^{g})$ to the values
reported in~(\ref{fit1param})-(\ref{fit9param}), we find minima with 
$\chi_{tot}^{2}\sim 500,$ which are clearly unacceptable.
This is a further confirmation of the incompatibility of results in the
scalar mesonic sector. If we attempt to reproduce the width of 
$K_0^{\ast}(1430)$, the other results are off. As noted above, this problem
arises in other models as well.

In~\cite{Eidelman:2004wy} only the $K_0^{\ast }\rightarrow K\pi $ mode is
quoted in the list of the decay modes. Our suggestion is that a strong
coupling to the scalar mesons below $1$ GeV, $K_0^{\ast}(800)$ and $\sigma$
takes place. Both states are very broad and could eventually influence the
decay strengths of $K_0^{\ast}(1430).$ The theoretical description of such a
phenomenon is however beyond the goals of the present tree-level study.

\section{Phenomenology involving the direct glueball decay mechanism}

\label{phen}

\subsection{Flavor symmetry limit}

The analysis of the previous section did not reveal a phenomenological need
to include the direct two-pseudoscalar decay of the glueball component. In
this section we analyze this additional mechanism when including it in the
fit. We start first by including the interaction term proportional to 
$c_{d}^{g}$, while neglecting the flavor symmetry breaking contribution 
$c_{m}^{g}.$

The ratio $c_{d}^{g}/c_{d}^{s}$ is a measure of the direct glueball decay
strength. The analogous quantity used in~\cite{Close:2001ga} is, because of
the different normalization, related as $c_{d}^{g}/c_{d}^{s} 
\longleftrightarrow \sqrt{3/2} \, r_{2}$. For the different solutions in the 
fit of~\cite{Close:2001ga} the parameter $r_{2}$ varies between $1$ and $5.$
Note that the case $r_{2}=1$ corresponds to a direct glueball decay strength
into two pseudoscalars $\sim 1.22$ larger than the quarkonium strength. 
In~\cite{Giacosa:2005qr} the bare masses have been kept fixed when the glueball
decay parameters have been introduced. Here we release this constraint by
leaving the bare masses free. In the fit we again find two different
solutions, which correspond to the ones (I and II) analyzed in the previous
section.

The $\chi^2$ minimum corresponding to solution I but now with inclusion of 
$c_{d}^{g}$ in the fit is: 
\begin{eqnarray}
\hspace*{-0.75cm}M_{N} &=&1.416~\text{\textrm{GeV}}, 
~M_{G}=1.493~\mathrm{GeV},~M_{S}=1.694~\mathrm{GeV}\,,  \nonumber \\
f &=&0.075~\mathrm{GeV}^{2},~\varepsilon =0.241~\mathrm{GeV}^{2},
~c_{d}^{s}=8.73~\mathrm{MeV}\,,  \nonumber \\
c_{m}^{s} &=&1.48~\mathrm{MeV}, ~c_{d}^{g}=-0.94~\mathrm{MeV};
~\chi_{tot}^{2}=28.70\,.
\end{eqnarray}
A comparison with the previous fit~(\ref{fitparam1}) shows that the
parameters practically did not change: with $c_{d}^{g}=-0.94$~MeV the direct
glueball decay is suppressed, resulting in $\left\vert
c_{d}^{g}/c_{d}^{s}\right\vert =0.11$ $\ll 1\,.$ Also, the total $\chi ^{2}$
is only slightly smaller than in~(\ref{fitparam1}).

The minimum corresponding to solution II reads: 
\begin{eqnarray}
\hspace*{-0.75cm}M_{N} &=&1.401~\mathrm{GeV},~M_{G}=1.461~\mathrm{GeV},
~M_{S}=1.609~\mathrm{GeV}\,,  \nonumber \\
f &=&0.376~\mathrm{GeV}^{2},~\varepsilon =-0.082~\mathrm{GeV}^{2},~
c_{d}^{s}=7.63~\mathrm{MeV}\,,  \nonumber \\
c_{m}^{s} &=&7.09~\mathrm{MeV}, ~c_{d}^{g}=1.82~\mathrm{MeV};
~\chi_{tot}^{2}=22.87\,.
\end{eqnarray}
A comparison with (\ref{fitparam2}) shows a clear similarity, although the
bare masses are somewhat shifted. Again, direct glueball decay is suppressed
with $\left\vert c_{d}^{g}/c_{d}^{s}\right\vert =0.24.$

The inclusion of $c_{d}^{g}$ in the fit does not lead to a drastic change of
the previous fit parameters. The bare glueball decay strength is strongly
suppressed with respect to the quarkonium one, thus in agreement with the
analysis of the previous section and with large $N_{c}$ considerations.

\subsection{Flavor symmetry breaking and lattice results}

Inclusion of the flavor symmetry breaking term in direct glueball decay,
that is the term with $c_{m}^{g}$, results in nine free parameters (the set
of~(\ref{param}) and $c_{d}^{g},$ $c_{m}^{g}$). A direct fit of these
parameters to the data generates various minima, which are not well
pronounced. For example, solutions are found where the flavor symmetry
breaking term in the glueball sector $c_{m}^{g}$ is exceedingly large,
dominating the decay mechanism. Instead, to study the effect of flavor
symmetry breaking in glueball decay we resort to a first lattice 
study~\cite{Sexton:1996ed} to fix the decay parameters $c_{d}^{g}$ 
and $c_{m}^{g}.$

In~\cite{Sexton:1996ed} a full two-pseudoscalar decay width of the scalar
glueball with about $100$ MeV is deduced. The corresponding mass of $1.7$
GeV led the authors of~\cite{Sexton:1996ed} to interpret the resonance 
$f_{0}(1710)$ as mainly gluonic. However, in the cited lattice analysis it is
not clear if the glueball decay mechanism occurs partially by mixing with
scalar quarkonia (here parametrized by $f$), or by direct decay
(parametrized by $c_{d}^{g}$ and $c_{m}^{g}$). In~\cite{Burakovsky:1998zg} a
scenario is studied, where the lattice results of~\cite{Sexton:1996ed} are
explained by mixing only (hence $c_{d}^{g}$ and $c_{m}^{g}$ are set to
zero). The physical state $f_{0}(1710)$ is mainly gluonic, but because of
mixing, it acquires a large $\overline{s}s$ amount, which it turn explains
the decay pattern. The corresponding mixing matrix is then similar to the
results of~\cite{Lee:1999kv} (and to some aspects of our solution II).

In~\cite{Lee:1999kv} it is stated that the amplitudes deduced 
in~\cite{Sexton:1996ed} probably include significant contributions from mixing 
of the scalar glueball with quarkonium, although not proven. If this is the
case, only the mixing mechanism contributes to glueball decay and then we
are back to the previous solutions, where the constants $c_{d}^{g}$ and 
$c_{m}^{g}$ are negligible. Here we also intend to investigate the opposite
case, where the decay couplings calculated in~\cite{Sexton:1996ed} arise
from direct glueball decay.

The decay widths of the glueball can be expressed as~\cite{Sexton:1996ed}
(see also~\cite{Burakovsky:1998zg}): 
\begin{eqnarray}
\Gamma _{G\rightarrow \pi \pi } &=&
\frac{3 \, \lambda ^{1/2}(M_{G}^{2},M_{\pi}^{2},M_{\pi }^{2})}
{32 \, \pi \, M_{G}^{3}}\{y_{G\pi \pi } \, M_{\rho }\}^{2}, \nonumber \\
\Gamma _{G\rightarrow \overline{K}K} &=&
\frac{\lambda^{1/2}(M_{G}^{2},M_{K}^{2},M_{K}^{2})}{8 \, \pi \, M_{G}^{3}}
\{y_{G\overline{K}K} \, M_{\rho }\}^{2},  \nonumber \\
\Gamma _{G\rightarrow \eta \eta } &=&\frac{\lambda ^{1/2}(M_{i}^{2},
M_{\eta}^{2},M_{\eta }^{2})}{32\pi M_{G}^{3}}\{y_{G\eta \eta }M_{\rho }\}^{2},
\label{latticedec}
\end{eqnarray} 
where $M_{\rho }=775$ MeV is the $\rho $ mass. The lattice results for the
decay constants $y_{G\pi \pi }$, $y_{G\overline{K}K}$ and $y_{G\eta \eta }$
are: 
\begin{equation}
y_{G\pi \pi }=0.834_{-0.579}^{+0.603},\text{ }y_{G\overline{K}K}
=2.654_{-0.402}^{+0.372},\text{ }y_{G\eta \eta }=3.099_{-0.423}^{+0.364}.
\end{equation} 
Note that in the flavor-symmetry limit we would expect 
$y_{G\pi \pi }:y_{G\overline{K}K}:y_{G\eta \eta }=1:1:1$. 
Although the errors are large, the
lattice results show a sizable deviation from this limit. As already noted,
it is however not clear, if and to what extent mixing with quarkonia is
included in these amplitudes. Interpreting the lattice results in the
context of the direct glueball decay mechanism implies a large symmetry
violation parameter $c_{m}^{g}.$

The corresponding decays can be derived from the expressions of Appendix B
by setting the glueball-quarkonia mixing to zero, that is by considering the
scalar-isoscalar decay for $M_{i}=M_{G}$ and $B_{iN}=B_{iS}=0$ and, of
course, $B_{iG}=1.$ The explicit expressions for the $\pi \pi $ and 
$\overline{K}K$ modes of the direct glueball decay are: 
\begin{eqnarray}
\Gamma _{G\rightarrow \pi \pi } &=&
\frac{3 \, \lambda ^{1/2}(M_{G}^{2},M_{\pi}^{2},M_{\pi }^{2})}
{32\pi M_{G}^{3}} \, \biggl\{ \frac{2}{\sqrt{3}F^{2}} 
\left( \left[
M_{G}^{2}-2M_{\pi }^{2}\right] c_{d}^{g}+2M_{\pi }^{2}c_{m}^{g}\right)
\biggr\}^{2},  \nonumber \\
\Gamma _{G\rightarrow \overline{K}K} &=&
\frac{\lambda^{1/2}(M_{G}^{2},M_{K}^{2},M_{K}^{2})}{8\pi M_{G}^{3}}
\biggl\{\frac{2}{\sqrt{3}F^{2}}\left( \left[ M_{G}^{2}-2M_{K}^{2}\right] 
c_{d}^{g}+2M_{K}^{2}c_{m}^{g}\right) \biggr\}^{2}.  \label{bareg}
\end{eqnarray} 
Matching the expressions for the $\pi \pi $ and $\overline{K}K$ decay modes
of (\ref{bareg}) to (\ref{latticedec}) using $M_{G}=1.7$ GeV 
(as in~\cite{Sexton:1996ed}) we obtain for the decay constants: 
\begin{equation}
c_{d}^{g}=1.34~\text{MeV ~ and ~ }c_{m}^{g}=24.6\;\text{MeV}.
\label{direct}
\end{equation} 
For a bare glueball mass of $M_{G}=1.5$ GeV we find rather similar values of 
$c_{d}^{g}=1.72$ MeV and $c_{m}^{g}=25$ MeV; hence we have a rather
slight dependence on $M_{G}$ within a reasonable range of values. In the
following we take the values evaluated at $M_{G}=1.7$ GeV.

Using the values of (\ref{direct}) we can also determine the $\eta \eta $
decay amplitude. Compared to the lattice result of 
$y_{G\eta \eta }~ M_{\rho }~=~2.40_{-0.33}^{+0.28}$~MeV 
we get the value of $2.025$ GeV. The
corresponding decay widths for the bare glueball are (for $M_{G}=1.7$~GeV): 
\begin{eqnarray}
\Gamma _{G\rightarrow \pi \pi } &=&7.23\text{ MeV, }\Gamma _{G\rightarrow 
\overline{K}K}=80.61\text{ MeV}, \\
\Gamma _{G\rightarrow \eta \eta } &=&18.35\text{ MeV, }\Gamma _{G\rightarrow
\eta \eta ^{\prime }}=11.73\text{ MeV}.
\end{eqnarray}%
For a lower value of mass $M_{G}$ the only significantly affected mode is 
$\eta \eta ^{\prime }$, since threshold effects become important (note that
the $\eta \eta ^{\prime }$ mode is entirely generated by the flavor symmetry
breaking term proportional to $c_{m}^{g}$).

With the direct glueball decay including flavor symmetry violation fixed 
by~(\ref{direct}), we now perform a fit with the remaining free parameters 
$M_{N},$ $M_{G},$ $M_{S},$ $f,$ $\varepsilon ,$ $c_{d}^{s},$ $c_{m}^{s}.$ 
Two solutions (III and IV) are obtained, which correspond to the bare level
orderings $M_{N}<M_{G}<M_{S}$ and $M_{N}<M_{G}<M_{S}$, which we analyze in
the following. Although the direct glueball decay dominates in these cases,
the two solutions have similarities to the ones discussed in detail in the
previous section.

\subsection{Third solution and implications}

Solution III is obtained for the set of parameters: 
\begin{eqnarray}
M_{N} &=&1.359~\mathrm{GeV},~M_{G}=1.435~\mathrm{GeV},
~M_{S}=1.686~\mathrm{GeV}\,,  \nonumber \\
f &=&0.212~\mathrm{GeV}^{2},~\varepsilon =0.277~\mathrm{GeV}^{2},  
\nonumber\\
~c_{d}^{s} &=&8.28~\mathrm{MeV},~c_{m}^{s}=7.21~\mathrm{MeV};
~\chi_{tot}^{2}=21.56\, ,
\end{eqnarray}
where the fit results are listed in Table 7.
The mixing matrix is similar to the one of solution I and explicitly reads:
\begin{center}
\begin{equation}
\,\left( 
\begin{array}{l}
\left\vert f_{1}\right\rangle \equiv \left\vert f_{0}(1370)\right\rangle \\ 
\left\vert f_{2}\right\rangle \equiv \left\vert f_{0}(1500)\right\rangle \\ 
\left\vert f_{3}\right\rangle \equiv \left\vert f_{0}(1710)\right\rangle 
\end{array}
\right) =\left( 
\begin{array}{lll}
0.79 & 0.56 & 0.26 \\ 
-0.58 & 0.81 & 0.02 \\ 
-0.20 & -0.16 & 0.97 
\end{array} 
\right) \left( 
\begin{array}{l}
\left\vert N\right\rangle \equiv \left\vert \bar{n}n\right\rangle \\ 
\left\vert G\right\rangle \equiv \left\vert gg\right\rangle \\ 
\left\vert S\right\rangle \equiv \left\vert \bar{s}s\right\rangle 
\end{array} 
\right) .
\end{equation}
\end{center}
The decay rates for $f_{1}=f_{0}(1370)$ and $f_{3}=f_{0}(1710)$ as compared
to the WA102 data are summarized in Tables 8 and 9.

The two-photon decay ratios resulting from solution III read: 
\begin{equation}
\Gamma _{f_{1}\rightarrow 2\gamma }:\Gamma _{f_{2}\rightarrow 2\gamma
}:\Gamma _{f_{3}\rightarrow 2\gamma }:\Gamma _{a_{0}^{0}\rightarrow 2\gamma
}=1:1.018:0.025:0.494.
\end{equation}

The results of the fit summarized in Table 7 are acceptable, apart from the
already discussed underestimate of the $K^{\ast }$ decay width. The
corresponding prediction for the WA102 data on $f_{0}(1370)$ has problems:
the predicted ratio $\Gamma _{f_{1}\rightarrow \bar{K}K}/\Gamma
_{f_{1}\rightarrow \pi \pi }$ is larger than unity and the full
two-pseudoscalar decay width is very large (we refer to the discussion of
solution I on the issue of the latter point). For the state $f_0(1710)$ we
obtain a large ratio $\Gamma _{f_{3}\rightarrow 
\eta \eta ^{\prime }}/\Gamma_{f_{3}\rightarrow \pi \pi }$ , 
again as in solution I, in contrast to the
WA102 result. For the two-photon decays we have a large ratio 
$\Gamma_{f_{2}\rightarrow 2\gamma }/\Gamma _{f_{1}\rightarrow 2\gamma }$.

\begin{center}
\textbf{Table 7.} Fitted mass and decay properties of scalar mesons.

\vspace*{0.5cm} 
\begin{tabular}{|l|l|l|l|}
\hline
Quantity & Exp & Theory & $\chi _{i}^{2}$ \\ \hline
$M_{f_{1}}$ \thinspace (\textrm{MeV}) & $1350$ $\pm 150$ & $1242$ & $0.519$
\\ 
$M_{f_{2}}$ \thinspace (\textrm{MeV}) & $1507\pm 5$ & $1507$ & $0.003$ \\ 
$M_{f_{3}}$ \thinspace (\textrm{MeV}) & $1714\pm 5$ & $1714$ & $0.011$ \\ 
$\Gamma _{f_{2}\rightarrow \pi \pi }$ \thinspace \thinspace (\textrm{MeV}) & 
$38.0\pm 4.6$ & $38.50$ & $0.010$ \\ 
$\Gamma _{f_{2}\rightarrow \overline{K}K}$\thinspace\thinspace (\textrm{MeV})& 
$9.4\pm 1.7$ & $10.38$ & $0.332$ \\ 
$\Gamma _{f_{2}\rightarrow \eta \eta }$ \thinspace (\textrm{MeV}) & $5.6\pm
1.3$ & $3.65$ & $2.252$ \\ 
$\Gamma _{f_{3}\rightarrow \pi \pi }/\Gamma_{f_{3}\rightarrow\overline{K}K}$ & 
$0.20\pm 0.06$ & $0.197$ & $0.002$ \\ 
$\Gamma _{f_{3}\rightarrow \eta \eta }/\Gamma_{f_{3}\rightarrow \overline{K}K}$
& $0.48\pm 0.15$ & $0.314$ & $1.221$ \\ 
$\Gamma _{a_{0}\rightarrow \bar{K}K}/\Gamma_{a_{0}\rightarrow \pi \eta }$ & 
$0.88\pm 0.23$ & $1.079$ & $0.745$ \\ 
$\Gamma _{a_{0}\rightarrow \pi \eta ^{\prime }}/\Gamma_{a_{0}\rightarrow
\pi \eta }$ & $0.35\pm 0.16$ & $0.291$ & $0.134$ \\ 
$\Gamma _{K_{0}^{\ast }\rightarrow K\pi }$ \thinspace (\textrm{MeV}) & 
$273\pm 51$ & $71.02$ & $16.221$ \\ 
$\left( \Gamma _{f_{3}}\right) _{2P}$ \thinspace (\textrm{MeV}) & $140\pm 10$
& $143.3$ & $0.109$ \\ 
$\chi _{tot}^{2}$ & - & - & $21.560$ \\ \hline
\end{tabular}
\end{center}

\begin{center}
\textbf{Table 8.} Decays of $f_{1}=f_{0}(1370)$.

\vspace*{0.5cm} 
\begin{tabular}{|l|l|l|}
\hline
Quantity & Exp (WA102) & Theory \\ \hline
$\Gamma _{f_{1}\rightarrow \overline{K}K}/
\Gamma _{f_{1}\rightarrow \pi \pi }$ & 
$~0.46\pm 0.19$ & $1.10$ \\ 
$\Gamma _{f_{1}\rightarrow \eta \eta }/\Gamma _{f_{1}\rightarrow \pi \pi }$
& $0.16\pm 0.07$ & $0.17$ \\ 
$\left( \Gamma _{f_{1}}\right) _{2P}$ (\textrm{MeV}) & \textquotedblright
small\textquotedblright & $193.5$ \\[1mm] \hline
\end{tabular}

\bigskip

\textbf{Table 9.} Decays of $f_{3}=f_{0}(1710)$.

\vspace*{0.5cm} 
\begin{tabular}{|l|l|l|}
\hline
Quantity & Exp (WA102) & Theory \\ \hline
$\Gamma _{f_{3}\rightarrow \overline{K}K}/
\Gamma _{f_{3}\rightarrow \pi \pi }$ & 
$5.0\pm 0.7$ & 5.08 \\ 
$\Gamma _{f_{3}\rightarrow \eta \eta }/\Gamma _{f_{3}\rightarrow \pi \pi }$
& $2.4\pm 0.6$ & 1.59 \\ 
$\Gamma _{f_{3}\rightarrow \eta \eta ^{\prime }}/\Gamma _{f_{3}\rightarrow
\pi \pi }$ & $<0.18$ & 2.01 \\ 
$\left( \Gamma _{f_{3}}\right) _{2P}$ (\textrm{MeV}) & \textquotedblright
dominant\textquotedblright & 143.3 \\[2mm] \hline
\end{tabular}
\end{center}

\subsection{Fourth solution and implications}

The solution IV corresponds to an inverted bare level ordering and a small
glueball-quarkonia mixing with 
\begin{eqnarray}
M_{N} &=&1.392~\mathrm{GeV},~M_{G}=1.712~\mathrm{GeV},
~M_{S}=1.452~\mathrm{GeV}\,,  \nonumber \\
f &=&-0.050~\mathrm{GeV}^{2},~\varepsilon =0.232~\mathrm{GeV}^{2},  
\nonumber\\
~c_{d}^{s} &=&6.66~\mathrm{MeV},~c_{m}^{s}=5.84~\mathrm{MeV};
~\chi_{tot}^{2}=26.330\,.
\end{eqnarray}

\begin{center}
\textbf{Table 10.} Fitted mass and decay properties of scalar mesons.

\vspace*{0.5cm} 
\begin{tabular}{|l|l|l|l|}
\hline
Quantity & Exp & Theory & $\chi _{i}^{2}$ \\ \hline
$M_{f_{1}}$ \thinspace (\textrm{MeV}) & $1350$ $\pm 150$ & $1330$ & $0.017$
\\ 
$M_{f_{2}}$ \thinspace (\textrm{MeV}) & $1507\pm 5$ & $1507$ & $0.002$ \\ 
$M_{f_{3}}$ \thinspace (\textrm{MeV}) & $1714\pm 5$ & $1714$ & $\sim 0$ \\ 
$\Gamma_{f_{2}\rightarrow \pi \pi }$ \thinspace \thinspace (\textrm{MeV}) & 
$38.0\pm 4.6$ & $39.84$ & $0.135$ \\ 
$\Gamma_{f_{2}\rightarrow \overline{K}K}$ 
\thinspace \thinspace (\textrm{MeV}) & 
$9.4\pm 1.7$ & $10.21$ & $0.230$ \\ 
$\Gamma_{f_{2}\rightarrow \eta \eta }$ \thinspace (\textrm{MeV}) & $5.6\pm
1.3$ & $4.89$ & $0.297$ \\ 
$\Gamma_{f_{3}\rightarrow \pi\pi}/\Gamma_{f_{3}\rightarrow\overline{K}K}$ & 
$0.20\pm 0.06$ & $0.129$ & $1.416$ \\ 
$\Gamma_{f_{3}\rightarrow \eta \eta }/\Gamma_{f_{3}\rightarrow\overline{K}K}$
& $0.48\pm 0.15$ & $0.227$ & $2.842$ \\ 
$\Gamma _{a_{0}\rightarrow \overline{K}K}/\Gamma_{a_{0}\rightarrow \pi\eta}$ & 
$0.88\pm 0.23$ & $1.086$ & $0.863$ \\ 
$\Gamma_{a_{0}\rightarrow \pi \eta ^{\prime }}/\Gamma_{a_{0}\rightarrow
\pi \eta }$ & $0.35\pm 0.16$ & $0.291$ & $0.134$ \\ 
$\Gamma_{K_{0}^{\ast }\rightarrow K\pi }$ \thinspace (\textrm{MeV}) & 
$273\pm 51$ & $45.05$ & $20.365$ \\ 
$\left( \Gamma _{f_{3}}\right) _{2P}$ \thinspace (\textrm{MeV}) & $140\pm 10$
& $142.95$ & $0.087$ \\ 
$\chi_{tot}^{2}$ & - & - & $26.330$ \\ \hline
\end{tabular}
\end{center}

Although the fit results given in Table 10 are acceptable, the smallness of
the glueball-quarkonia mixing is clear contrast to other phenomenological
studies~\cite{Close:2001ga}-\cite{Giacosa:2005qr} and lattice 
result~\cite{Lee:1999kv,Sexton:1996ed}.

The mixing matrix reads:

\begin{center}
\begin{equation}
\,\left( 
\begin{array}{l}
\left\vert f_{1}\right\rangle \equiv \left\vert f_{0}(1370)\right\rangle \\ 
\left\vert f_{2}\right\rangle \equiv \left\vert f_{0}(1500)\right\rangle \\ 
\left\vert f_{3}\right\rangle \equiv \left\vert f_{0}(1710)\right\rangle 
\end{array}
\right) =\left( 
\begin{array}{lll}
0.82 & -0.07 & 0.57 \\ 
-0.57 & \sim 0 & 0.82 \\ 
-0.06 & 0.99 & 0.04 
\end{array} 
\right) \left( 
\begin{array}{l}
\left\vert N\right\rangle \equiv \left\vert \bar{n}n\right\rangle \\ 
\left\vert G\right\rangle \equiv \left\vert gg\right\rangle \\ 
\left\vert S\right\rangle \equiv \left\vert \bar{s}s\right\rangle 
\end{array}
\right) .
\end{equation}
\end{center}

In this solution the state $f_{0}(1710)$ is very close to a pure gluonic
configuration, which is traced to the small mixing parameter $f$. The states 
$f_{0}(1370)$ and $f_{0}(1500)$ are in turn dominated by the quarkonia
components, but with strong mixing between $\overline{n}n$ and 
$\overline{s}s$.
The decay rates for $f_{1}=f_{0}(1370)$ and $f_{3}=f_{0}(1710)$ as compared
to the WA102 data are listed in Tables 11 and 12.

\begin{center}
\textbf{Table 11.} Decays of $f_{1}=f_{0}(1370)$.

\vspace*{0.5cm} 
\begin{tabular}{|l|l|l|}
\hline
Quantity & Exp (WA102) & Theory \\ \hline
$\Gamma_{f_{1}\rightarrow \overline{K}K}/\Gamma_{f_{1}\rightarrow \pi\pi }$ & 
$~0.46\pm 0.19$ & $0.73$ \\ 
$\Gamma_{f_{1}\rightarrow \eta \eta }/\Gamma _{f_{1}\rightarrow \pi \pi }$
& $0.16\pm 0.07$ & $0.13$ \\ 
$\left( \Gamma_{f_{1}}\right) _{2P}$ (\textrm{MeV}) & \textquotedblright
small\textquotedblright & $99.47$ \\[1mm] \hline
\end{tabular}

\textbf{Table 12.} Decays of $f_{3}=f_{0}(1710)$.

\vspace*{0.5cm} 
\begin{tabular}{|l|l|l|}
\hline
Quantity & Exp (WA102) & Theory \\ \hline
$\Gamma _{f_{3}\rightarrow \overline{K}K}/
\Gamma _{f_{3}\rightarrow \pi \pi }$ & 
$5.0\pm 0.7$ & $7.78$ \\ 
$\Gamma _{f_{3}\rightarrow \eta \eta }/\Gamma _{f_{3}\rightarrow \pi \pi }$
& $2.4\pm 0.6$ & $1.76$ \\ 
$\Gamma _{f_{3}\rightarrow \eta \eta ^{\prime }}/\Gamma _{f_{3}\rightarrow
\pi \pi }$ & $<0.18$ & $1.00$ \\ 
$\left( \Gamma _{f_{3}}\right) _{2P}$ (\textrm{MeV}) & \textquotedblright
dominant\textquotedblright & $142.95$ \\[2mm] \hline
\end{tabular}
\end{center}

Finally, the two-photon decay ratios resulting from solution IV read: 
\begin{equation}
\Gamma _{f_{1}\rightarrow 2\gamma }:\Gamma _{f_{2}\rightarrow 2\gamma
}:\Gamma _{f_{3}\rightarrow 2\gamma }:\Gamma _{a_{0}^{0}\rightarrow 2\gamma
}=1:0.273:0.008:0.382.
\end{equation}

\subsection{Discussion of solutions III and IV}

The study of the previous subsections shows that a direct glueball decay
with large violation of flavor symmetry is feasible. This feature is based
on an interpretation of first lattice results, which corresponds to the
limiting case of a direct glueball decay as contained in the parameters 
$c_{d}^{g}$ and $c_{m}^{g})$. This interpretation is in contrast to the
phenomenological study of \cite{Burakovsky:1998zg} and to the comments given
in~\cite{Sexton:1996ed}. Furthermore, such a large value for the flavor
symmetry breaking parameter $c_{m}^{g}$ is not in agreement with large $N_{c}
$ arguments. In any case, considering that the interpretation of lattice
results is not unique, from a phenomenological point of view it is
interesting to analyze the situation, where a strong, flavor symmetry
violating glueball decay is present.

A sizable direct glueball decay is, as already discussed, the result of the
phenomenology given in~\cite{Close:2001ga}. However, in~\cite{Close:2001ga}
the direct glueball decay pattern is flavor blind, which would correspond to
a large $c_{d}^{g}$ but to a suppressed value for $c_{m}^{g}$. In this sense
the solutions III and IV differ from the analysis of \cite{Close:2001ga}.

Care should also be taken when considering the two-photon decay in this
scheme: if the two-pseudoscalar amplitudes are sizable, the same can also be
expected for the transitions into two vector mesons (although not studied
here). Invoking vector meson dominance sizable corrections to the two-photon
final state are expected. The use of the limit $c_{e}^{g}=0$ in the present
case is therefore questionable.

\section{Conclusions}

\label{concl}

In this paper we analyzed the two-pseudoscalar and the two-photon decays of
the scalar states between $1$-$2$ GeV in the framework of a chiral
Lagrangian, where the glueball has been included as a flavor-blind composite
mesonic field with independent couplings to pseudoscalar fields.

In a first step we have set the glueball-pseudoscalar couplings to zero and
performed a fit to the accepted averages of PDG2004~\cite{Eidelman:2004wy}.
We find two possible solutions (I and II), which, apart from the
underestimate of the $K_{0}^{\ast }\rightarrow K\pi $ mode, show good
agreement with the data. The solutions I and II differ in the bare isoscalar
masses, in the mixing matrix (in scheme I) the state $f_{0}(1500)$ 
has the largest
gluonic amount, while in scheme II the state $f_{0}(1710)$ has the main 
gluonic component), in some predictions concerning other decay modes, 
which have been compared to the experimental results 
of~\cite{Barberis:2000cd}. From a phenomenological point of view, 
there is no striking hint for a direct glueball-pseudoscalar coupling.

We then enlarged our analysis by including the direct glueball decay
parameter ($c_{d}^{g}$) in the flavor symmetry limit in the fit. A small
value for this parameter is obtained, which in turn confirms the suppression
of the direct glueball decay in agreement with large $N_{c}$ arguments and
with~\cite{Amsler:1995tu,Strohmeier-Presicek:1999yv}, but contrary to the
study of~\cite{Close:2001ga}. In a last step we also included the second
glueball decay parameter ($c_{m}^{g}$) in the fit, which involves flavor
symmetry breaking in the direct glueball decay. The minima in the fit are
less pronounced, therefore we utilized the lattice results 
of~\cite{Sexton:1996ed} to determine the direct glueball decay 
parameters $c_{d}^{g}$
and $c_{m}^{g}$. The lattice data are interpreted such as they are matched
by the direct glueball decay mechanism, although this procedure might be a
limiting case. The resulting fits also generate a good description of the
data, where either the $f_0(1500)$ or the $f_0(1710)$ contain a dominant
glueball component. These solutions however should at present taken with
some care and require further input either from an enlarged, reliable
database or lattice constraints. For this reason our preferred solutions 
are I and II presented in section III.

Although the presence of a strong direct glueball decay cannot be verified
directly, the presence of a sizable glueball-quarkonia mixing is essential
to be in accord with the data. The magnitude is different in the two
proposed solutions (smaller in I, largest in II), but is in line with other
models and in magnitude qualitatively consistent with lattice 
results~\cite{Lee:1999kv,McNeile:2000xx}.

The starting point of the Lagrangian has been outlined 
in~\cite{Cirigliano:2003yq}. In the present work we added the glueball degree 
of freedom, both for mixing and decays. As a result of the fit, we also
considered deviations from the large $N_{c}$ limit. We find that large 
$N_{c} $ arguments are still useful as a guideline in the scalar sector as
well. Also a small deviation from the GMO octet mass relation is found.

A direct isoscalar $\overline{n}n$-$\overline{s}s$ quarkonia mixing has been
introduced in the theoretical analysis. Instanton solutions of the QCD
vacuum are believed to generate strong flavor-mixing in both the isoscalar
and the scalar sectors. The presence of such mixing is established in the
pseudoscalar nonet, while it is still an open question in the
scalar-isoscalar mesonic sector. Our two proposed solutions differ in this
point: while in the first one this flavor mixing sensibly affects the
results, in the second solution it turns out to be negligible.

The problem of the $K_0^{\ast}\rightarrow K\pi $ mode has been discussed;
although no final statement to this puzzle can be said, we compared
predictions of various approach and attempted to highlight the difficulty in
the scalar sector.

Although many experimental results can be reproduced, and the presence of a
scalar glueball and its mixing with scalar quarkonia explains many features
of the scalar meson spectroscopy, further work, both theoretically and
experimentally, is needed to rule out some mixing scenarios in favour of
others.

\begin{acknowledgments}
This work was supported by the by the DFG under contracts FA67/25-3 and
GRK683. This research is also part of the EU Integrated Infrastructure
Initiative Hadron physics project under contract number RII3-CT-2004-506078
and President grant of Russia "Scientific Schools" No. 1743.2003.
\end{acknowledgments}

\newpage

\appendix 

\section{Aspects of the chiral Lagrangian}

\label{appendixA}

We use the following phase conventions for the pseudoscalar $P=\{\pi ^{\pm
},\pi ^{0},K^{\pm },K^{0},\overline{K^{0}},\eta ^{0},\eta ^{8}\}$ and scalar 
\newline
$S=\{a_{0}^{\pm },a_{0}^{0},K_{0}^{\ast \pm },K_{0}^{\ast 0},
\overline{K_0^{\ast 0}},S^{0},S^{8}\}$ meson fields 
(neglecting the mixing of the third and the eighth component): 
\begin{eqnarray}
&&\pi ^{\pm }=\frac{1}{\sqrt{2}}(P_{1}\mp iP_{2})\,,\hspace*{0.5cm}\pi
^{0}=P_{3}\,,\hspace*{0.5cm}K^{\pm }=\frac{1}{\sqrt{2}}(P_{4}\mp iP_{5})\,, 
\nonumber  \label{fields} \\
&&K^{0}=\frac{1}{\sqrt{2}}(P_{6}-iP_{7})\,,\hspace*{0.5cm}
\overline{K^0}=
\frac{1}{\sqrt{2}}(P_{6}+iP_{7})\,,\hspace*{0.5cm}\eta ^{0}=P_{0}\,,
\hspace*{0.5cm} \eta ^{8}=P_{8}\,,  \nonumber \\
&& \\
&&a_{0}^{\pm }=\frac{1}{\sqrt{2}}(S_{1}\mp iS_{2})\,,\hspace*{0.8cm} 
a_{0}^{0}=S_{3}\,,\hspace*{0.5cm}K_{0}^{\ast \pm }=\frac{1}{\sqrt{2}} 
(S_{4}\mp iS_{5})\,,  \nonumber \\
&&K_{0}^{\ast 0}=\frac{1}{\sqrt{2}}(S_{6}-iS_{7})\,,\hspace*{0.5cm}
\overline{K_0^{\ast 0}}=\frac{1}{\sqrt{2}}(S_{6}+iS_{7})\,.  \nonumber
\end{eqnarray}

\section{Two-body $s \to p_1 p_2$ and $s \to \gamma \gamma$ transitions 
(matrix elements and decay widths)}\label{spp_widths}

\subsection{\protect\medskip Scalar-isoscalar strong decays}

The strong decay widths of the scalar states are derived at tree-level from
the following term of the Lagrangian (\ref{L_eff}):

\begin{equation}
\mathcal{L}_{decay}^{strong}=c_{d}^{s}\,\left\langle \mathcal{S}\,u_{\mu}
\,u^{\mu }\,\right\rangle +\,c_{m}^{s}\left\langle \mathcal{S}\,\chi_{+}
\right\rangle \,\,+\frac{c_{d}^{g}}{\sqrt{3}}G\,\left\langle u_{\mu}
\,u^{\mu }\right\rangle \,+\,\frac{c_{m}^{g}}{\sqrt{3}}\,G\left\langle 
\chi_{+}\right\rangle .
\end{equation}
The decay expression for the scalar-isoscalar states $\left\vert
i\right\rangle $ with $i=f_{1},f_{2},f_{3}$ into $\pi \pi ,$ $\overline{K}K,$
$\eta \eta $ and $\eta \eta ^{\prime }$ are given by the following
expressions:
\begin{eqnarray}
\Gamma _{i\rightarrow \pi \pi } &=&\Gamma _{i\rightarrow \pi ^{+}\pi^{-}}
+\Gamma _{i\rightarrow \pi ^{0}\pi ^{0}}=\frac{3}{2}\,\,\Gamma
_{i\rightarrow \pi ^{+}\pi ^{-}}=\frac{3\,\lambda ^{1/2}(M_{i}^{2},M_{\pi
}^{2},M_{\pi }^{2})}{32\,\pi \,M_{i}^{3}}\,|M_{i\pi ^{+}\pi ^{-}}|^{2}\,, \\
&&  \nonumber \\
\Gamma _{i\rightarrow K\overline{K}} &=&\Gamma _{i\rightarrow K^{+}K^{-}}
+\Gamma_{i\rightarrow K^{0}\bar{K}^{0}}=2\,\,\Gamma _{i\rightarrow K^{+}K^{-}}= 
\frac{\lambda ^{1/2}(M_{i}^{2},M_{K}^{2},M_{K}^{2})}{8\,\pi \,M_{i}^{3}} 
\,|M_{iK^{+}K^{-}}|^{2}\,, \\
&&  \nonumber \\
\Gamma _{i\rightarrow \eta \eta } &=&\frac{\lambda ^{1/2}(M_{i}^{2},
M_{\eta}^{2},M_{\eta }^{2})}{32\,\pi \,M_{i}^{3}}\,|M_{i\eta \eta }|^{2}\,, \\
&&  \nonumber \\
\Gamma _{i\rightarrow \eta \eta ^{\prime }} &=&\frac{\lambda^{1/2}(M_{i}^{2},
M_{\eta }^{2},M_{\eta ^{\prime }}^{2})}{16\,\pi \,M_{i}^{3}}
\,|M_{i\eta \eta ^{\prime }}|^{2} 
\end{eqnarray} 
where $\lambda (x,y,z)$ is the K\"{a}llen triangle function: 
\begin{equation}
\lambda (x,y,z)=x^{2}+y^{2}+z^{2}-2xy-2yz-2xz \,. 
\end{equation}
The matrix elements $M_{i\pi ^{+}\pi ^{-}}$, $M_{iK^{+}K^{-}}$, 
$M_{i\eta\eta }$ and $M_{i\eta \eta ^{\prime }}$ are given by 
\begin{eqnarray}
\hspace*{-0.5cm}M_{i\rightarrow \pi ^{+}\pi ^{-}} &=&
- \frac{2\,B_{iN}}{F^{2}\, 
\sqrt{2}}\,\biggl\{[M_{i}^{2}\,-\,2\,M_{\pi }^{2}]
\,c_{d}^{s}\,+\,2\,M_{\pi}^{2}\,c_{m}^{s}\biggr\}\,
- \,\frac{2\,B_{iG}}{F^{2}\,\sqrt{3}}\biggl\{%
[M_{i}^{2}\,-\,2\,M_{\pi }^{2}]\,c_{d}^{g}\,+\,2\,M_{\pi }^{2}\,c_{m}^{g}%
\biggr\}\,, \\
\hspace*{-0.5cm} &&  \nonumber \\
\hspace*{-0.5cm}M_{i\rightarrow K^{+}K^{-}} &=& - 
\frac{B_{iN}+\sqrt{2}B_{iS}}{F^{2}\sqrt{2}}
\,\biggl\{[M_{i}^{2}\,-\,2\,M_{K}^{2}]\,c_{d}^{s}\,+\,2%
\,M_{K}^{2}\,c_{m}^{s}\biggr\}  \nonumber \\
&-&\frac{2\,B_{iG}}{F^{2}\,\sqrt{3}}\biggl\{[M_{i}^{2}\,-\,2\,M_{K}^{2}]%
\,c_{d}^{g}\,+\,2\,M_{K}^{2}\,c_{m}^{g}\biggr\}\,, \\
\hspace*{-.5cm} & &\nonumber\\
\hspace*{-.5cm} M_{i \rightarrow  \eta \eta} &=& 
- \frac{2 \,  c_{d}^{s}}{F^2 \, \sqrt{2}} \, [M_i^2- 2M_\eta^2 ] \, 
\biggl\{ B_{iN} \, \sin^2\delta_P 
\, + \, B_{iS} \, \cos^2\delta_P \, \sqrt{2} \biggr\} \nonumber \\
\hspace*{-.5cm} &-& 
\frac{4 \,  c_{m}^{s}}{F^2 \, \sqrt{2}} \, 
\biggl\{ M_\pi^{2} \, B_{iN} \, \sin^2\delta_P 
\, +  \, [2 \, M_K^2 \, - \, M_\pi^2] B_{iS} 
\,  \cos^2\delta_P \, \sqrt{2}\biggr\}  \\
\hspace*{-.5cm} &-& \frac{2 \, B_{iG}}{F^2 \, \sqrt{3}} \, 
\biggl\{ c_d^g \, [M_i^2 \, - \, 2 M_\eta^2] \, + \, 
2 \, c_m^g \, 
\biggl( M_\pi^2 \, + \, 2 \, [M_K^2 - M_\pi^2] \, \cos^2\delta_P 
\biggr) \biggr\} \,, \nonumber \\
\hspace*{-.5cm} & &\nonumber\\  
M_{i \rightarrow \eta \eta^\prime} &=& \frac{\sin 2\delta_P}{F^2 \sqrt{2}} 
\biggl\{ c_d^s [M_i^2-M_\eta^2-M_{\eta^\prime}^2] 
[ B_{iN} \, - \, B_{iS} \, \sqrt{2} ] \nonumber\\
\hspace*{-.5cm} 
&+& 2 c_{m}^{s} [ M_\pi^2 \, B_{iN} \, + \, 
(M_\pi^2 - 2 M_K^2) \, B_{iS} \, \sqrt{2} ] 
- 4 \sqrt{\frac{2}{3}} \, c_m^g \,  
B_{iG} [M_K^2 - M_\pi^2] \biggr\} \,, 
\end{eqnarray} 
where $\delta_P = \theta_P - \theta_P^I$ and 
$\theta_P^I$ is the ideal mixing angle with 
$\sin\theta_P^I = 1/\sqrt{3}$; the quantities $B_{ij}$ are the 
elements of mixing matrix relating physical states $i=f_1,f_2,f_3$ and 
bare states $j=N,G,S$ (see definitions in Eqs.~\ref{S_08_to_NS}) 
and~(\ref{fi_to_NGS})). 

\subsection{Isovector and isodoublet strong decays}

\bigskip The decay rates for $a_{0}(1450)$ into $K\overline{K},$ $\pi \eta $
and $\pi \eta ^{\prime }$ are: 
\begin{eqnarray}
\Gamma _{a_{0}\rightarrow K\overline{K}} &=&\frac{1}{3}
\biggl[\Gamma_{a_{0}^{+}\rightarrow K^{+}\overline{K^{0}}}
+\Gamma _{a_{0}^{-}\rightarrow
K^{-}K^{0}}+\Gamma _{a_{0}^{0}\rightarrow K^{+}K^{-}}
+\Gamma_{a_{0}^{0}\rightarrow K^{0}\overline{K^{0}}}\biggr] \nonumber\\
&=&\,\Gamma _{a_{0}^{+}\rightarrow K^{+}\overline{K}^{0}}=
\frac{\lambda^{1/2}(M_{a_{0}}^{2},M_{K}^{2},M_{K}^{2})}{16\pi \,M_{a_{0}}^{3}}
\,|M_{a_{0}^{+}K^{+}\overline{K}^{0}}|^{2}\,, \\
&&  \nonumber \\
\Gamma _{a_{0}\rightarrow \pi \eta } &=&\frac{1}{3}
\biggl[\Gamma_{a_{0}^{+}\rightarrow \pi ^{+}\eta }+\Gamma _{a_{0}^{-}
\rightarrow \pi^{-}\eta }+\Gamma _{a_{0}^{0}\rightarrow \pi ^{0}\eta }\biggr] 
\nonumber\\
&=&\Gamma _{a_{0}^{+}\rightarrow \pi ^{+}\eta }=
\frac{\lambda^{1/2}(M_{a_{0}}^{2},M_{\pi }^{2},M_{\eta }^{2})}
{16\,\pi \,M_{a_{0}}^{3}} \,|M_{a_{0}^{+}\pi +\eta }|^{2}\,, \\
&&  \nonumber \\
\Gamma _{a_{0}\rightarrow \pi \eta ^{\prime }} &=&\frac{1}{3}
\biggl[\Gamma_{a_{0}^{+}\rightarrow \pi ^{+}\eta ^{\prime }}+
\Gamma_{a_{0}^{-}\rightarrow \pi ^{-}\eta ^{\prime }}+
\Gamma_{a_{0}^{0}\rightarrow \pi ^{0}\eta ^{\prime }}\biggr] \nonumber\\
&=&\Gamma _{a_{0}^{+}\rightarrow \pi ^{+}\eta ^{\prime }}=
\frac{\lambda^{1/2}(M_{a_{0}}^{2},M_{\pi }^{2},M_{\eta ^{\prime }}^{2})}
{16\,\pi\,M_{a_{0}}^{3}}\,|M_{a_{0}^{+}\pi +\eta ^{\prime }}|^{2}\,.
\end{eqnarray} 
The matrix elements $M_{a_{0}^{+}K^{+}\bar{K}^{0}}$, 
$M_{a_{0}^{+}\pi^{+}\eta }$ and $M_{a_{0}^{+}\pi ^{+}\eta ^{\prime }}$ 
are given by:

\begin{eqnarray}
M_{a_{0}^{+}K^{+}\bar{K}^{0}} &=&- \frac{1}{F^{2}}\,
\biggl([M_{a_{0}}^{2}-2M_{K}^{2}]c_{d}^{s}+2M_{K}^{2}\,c_{m}^{s}\biggr)\, \\
&&  \nonumber \\
M_{a_{0}^{+}\pi ^{+}\eta } &=&\frac{\sin \delta _{P} \, \sqrt{2}}{F^{2}}
\biggl([M_{a_{0}}^{2}-M_{\pi }^{2}-M_{\eta }^{2}]c_{d}^{s}+2M_{\pi }^{2}
c_{m}^{s} \biggr)\,, \\
&&  \nonumber \\
M_{a_{0}^{+}\pi ^{+}\eta ^{\prime }} &=&
- \frac{\cos \delta _{P} \, \sqrt{2}}{F^{2}}
\biggl([M_{a_{0}}^{2}-M_{\pi }^{2}-M_{\eta ^{\prime}}^{2}]c_{d}^{s}
+2M_{\pi }^{2}c_{m}^{s}\biggr)\,. 
\end{eqnarray}

The decay rates for $K_{0}^{\ast }(1430)$ into $K\pi $ and $K\eta $ are
(considering the isodoublet $\{K_{0}^{\ast +},K_{0}^{\ast 0}\}$): 
\begin{eqnarray}
\Gamma _{K_{0}^{\ast }\rightarrow K\pi } &=& 
\frac{1}{2}\biggl[\Gamma_{K_{0}^{\ast \,+}\rightarrow K^{0}\pi ^{+}}
+\Gamma _{K_{0}^{\ast\,+}\rightarrow K^{+}\pi ^{0}}
+\Gamma _{K_{0}^{\ast \,0}\rightarrow K^{0}\pi^{0}}+
\Gamma _{K_{0}^{\ast \,0}\rightarrow K^{+}\pi ^{-}}\biggr]  \nonumber\\
&=&\frac{3}{2}\,\,\Gamma _{K_{0}^{\ast \,+}\rightarrow K^{0}\pi ^{+}}=
\frac{3\,\lambda^{1/2}(M_{K_{0}^{\ast }}^{2},M_{K}^{2},M_{\pi }^{2})}
{32\,\pi\,M_{K_{0}^{\ast }}^{3}}\,|M_{K_{0}^{\ast \,+}K^{0}\pi ^{+}}|^{2}\,, \\
&&  \nonumber \\
\Gamma _{K_{0}^{\ast }\rightarrow K\eta } &=&\frac{1}{2}
\biggl[\Gamma_{K_{0}^{\ast \,+}\rightarrow K^{+}\eta}
+\Gamma _{K_{0}^{\ast\,0}\rightarrow K^{0}\eta}\biggr] \nonumber\\
&=&\Gamma _{K_{0}^{\ast \,+}\rightarrow K^{+}\eta}
=\frac{\lambda^{1/2}(M_{K_{0}^{\ast }}^{2},M_{K}^{2},M_{\eta }^{2})}
{16\,\pi\,M_{K_{0}^{\ast }}^{3}}\,|M_{K_{0}^{\ast \,+}K^{+}\eta }|^{2}\,,
\end{eqnarray} 
The matrix elements $M_{K_{0}^{\ast \,+}K^{0}\pi ^{+}}$ and 
$M_{K_{0}^{\ast\,+}K^{+}\eta }$ are given by 
\begin{eqnarray}
M_{K_{0}^{\ast \,+}K^{0}\pi ^{+}} &=& - \frac{1}{F^{2}}
\,\biggl([M_{K_{0}^{\ast }}^{2}-M_{\pi }^{2}-M_{K}^{2}]\,c_{d}^{s}
\, + \, [M_{\pi}^{2}+M_{K}^{2}]\,c_{m}^{s}\biggr)\,, \\
&&  \nonumber \\
M_{K_{0}^{\ast \,+}K^{+}\eta } &=&\frac{\cos \delta _{P}}{F^{2}}
\biggl([M_{K_{0}^{\ast }}^{2}-M_{K}^{2}-M_{\eta}^{2}]\,c_{d}^{s}
+[3M_{K}^{2}-M_{\pi }^{2}]\,c_{m}^{s}\biggr)  \nonumber \\
&&+\frac{\sin \delta _{P}}{F^{2}\,\sqrt{2}}\biggl([M_{K_{0}^{\ast}}^{2}
-M_{K}^{2}-M_{\eta }^{2}]\,c_{d}^{s}
+[M_{K}^{2}+M_{\pi}^{2}]\,c_{m}^{s}\biggr) \,.  
\end{eqnarray}


\begin{thebibliography}{99}

\bibitem{Eidelman:2004wy} S.~Eidelman \textit{et al.} 
[Particle Data Group Collaboration], 
Phys.\ Lett.\ B \textbf{592}, 1 (2004). 

\bibitem{Michael:2003ai} C.~Michael, 
arXiv:hep-lat/0302001; 
G.~S.~Bali, K.~Schilling, A.~Hulsebos, A.~C.~Irving, C.~Michael and
P.~W.~Stephenson [UKQCD Collaboration], 
Phys.\ Lett.\ B \textbf{309}, 378 (1993) [arXiv:hep-lat/9304012]; 
G.~S.~Bali \textit{et al.} [TXL Collaboration], 
Phys.\ Rev.\ D \textbf{62}, 054503 (2000) [arXiv:hep-lat/0003012]; 
C.~Morningstar and M.~J.~Peardon, 
AIP Conf.\ Proc.\ \textbf{688}, 220 (2004) [arXiv:nucl-th/0309068]. 

\bibitem{Amsler:2004ps} C.~Amsler and N.~A.~Tornqvist, 
Phys.\ Rept.\ \textbf{389}, 61 (2004). 

\bibitem{Close:2002zu} F.~E.~Close and N.~A.~Tornqvist, 
J.\ Phys.\ G \textbf{28}, R249 (2002) [arXiv:hep-ph/0204205]; 
E.~Klempt, 
arXiv:hep-ex/0101031; 
S.~F.~Tuan, 
arXiv:hep-ph/0303248; 
G.~V.~Efimov and M.~A.~Ivanov, ''The Quark Confinement Model of Hadrons'',
IOP Publishing, Bristol $\&$ Philadelphia, 1993. 

\bibitem{Narison:1996fm} S.~Narison,  
Nucl.\ Phys.\ B \textbf{509}, 312 (1998)  [arXiv:hep-ph/9612457].

\bibitem{Amsler:1995tu} C.~Amsler and F.~E.~Close, 
Phys.\ Lett.\ B \textbf{353}, 385 (1995) [arXiv:hep-ph/9505219]; 
C.~Amsler and F.~E.~Close, 
Phys.\ Rev.\ D \textbf{53}, 295 (1996) [arXiv:hep-ph/9507326]. 

\bibitem{Lee:1999kv} W.~J.~Lee and D.~Weingarten, 
Phys.\ Rev.\ D \textbf{61}, 014015 (2000) [arXiv:hep-lat/9910008]; 
D.~Weingarten, 
Nucl.\ Phys.\ Proc.\ Suppl.\ \textbf{53}, 232 (1997)
[arXiv:hep-lat/9608070]. 

\bibitem{Close:2001ga} F.~E.~Close and A.~Kirk, 
Eur.\ Phys.\ J.\ C \textbf{21}, 531 (2001) [arXiv:hep-ph/0103173]. 

\bibitem{Burakovsky:1998zg} L.~Burakovsky and P.~R.~Page, 
Phys.\ Rev.\ D \textbf{59}, 014022 (1999) [Erratum-ibid.\ D \textbf{59},
079902 (1999)] [arXiv:hep-ph/9807400]. 

\bibitem{Strohmeier-Presicek:1999yv} M.~Strohmeier-Presicek, T.~Gutsche,
R.~Vinh Mau and A.~Faessler, 
Phys.\ Rev.\ D \textbf{60}, 054010 (1999) [arXiv:hep-ph/9904461]. 

\bibitem{Ivanov:1985zw} M.~A.~Ivanov and R.~K.~Muradov, 
JETP Lett.\ \textbf{42}, 367 (1985); 
M.~Jaminon and B.~van den Bossche, 
Nucl.\ Phys.\ A \textbf{636}, 113 (1998) [arXiv:nucl-th/9712029]; 
J.~V.~Burdanov and G.~V.~Efimov, 
Phys.\ Rev.\ D \textbf{64}, 014001 (2001) [arXiv:hep-ph/0009027]; 
M.~K.~Volkov and V.~L.~Yudichev, 
Eur.\ Phys.\ J.\ A \textbf{10}, 223 (2001) [arXiv:hep-ph/0103003]. 

\bibitem{Giacosa:2004ug} F.~Giacosa, T.~Gutsche and A.~Faessler, 
Phys. Rev. C \textbf{71}, 025202 (2005) [arXiv:hep-ph/0408085]. 

\bibitem{Giacosa:2005qr}
F.~Giacosa, T.~Gutsche, V.~E.~Lyubovitskij and A.~Faessler, 
Phys.\ Lett.\ B {\bf 622}, 277 (2005) [arXiv:hep-ph/0504033]. 

\bibitem{Prelovsek:2004jp} S.~Prelovsek, C.~Dawson, T.~Izubuchi, K.~Orginos
and A.~Soni, 
Phys.\ Rev.\ D \textbf{70} (2004) 094503 [arXiv:hep-lat/0407037]. 

\bibitem{McNeile:2000xx} C.~McNeile and C.~Michael [UKQCD Collaboration], 
Phys.\ Rev.\ D \textbf{63} (2001) 114503 [arXiv:hep-lat/0010019]. 

\bibitem{Bali:2003tp} G.~S.~Bali, 
Eur.\ Phys.\ J.\ A \textbf{19} (2004) 1 [arXiv:hep-lat/0308015]. 

\bibitem{Alford:2000mm} M.~G.~Alford and R.~L.~Jaffe, 
Nucl.\ Phys.\ B \textbf{578} (2000) 367 [arXiv:hep-lat/0001023]. 

\bibitem{Pelaez:2003dy} J.~R.~Pelaez, 
Phys.\ Rev.\ Lett.\ \textbf{92} (2004) 102001 [arXiv:hep-ph/0309292]. 

\bibitem{Oller:1997ti} J.~A.~Oller and E.~Oset, 
Nucl.\ Phys.\ A \textbf{620} (1997) 438 [Erratum-ibid.\ A \textbf{652}
(1999) 407] [arXiv:hep-ph/9702314]. 

\bibitem{Cirigliano:2003yq} V.~Cirigliano, G.~Ecker, H.~Neufeld and A.~Pich, 
JHEP \textbf{0306} (2003) 012 [arXiv:hep-ph/0305311]. 

\bibitem{ChPT} 
S.~Weinberg, 
Physica A \textbf{96} (1979) 327; 
J.~Gasser and H.~Leutwyler, 
Annals Phys.\ \textbf{158}, 142 (1984); 
Nucl.\ Phys.\ B \textbf{250} (1985) 465. 
\bibitem{Ecker:1988te} G.~Ecker, J.~Gasser, A.~Pich and E.~de Rafael, 
Nucl.\ Phys.\ B \textbf{321}, 311 (1989); 
G.~Ecker, J.~Gasser, H.~Leutwyler, A.~Pich and E.~de Rafael, 
Phys.\ Lett.\ B \textbf{223}, 425 (1989). 

\bibitem{Klempt:1995ku} E.~Klempt, B.~C.~Metsch, C.~R.~Munz and H.~R.~Petry, 
Phys.\ Lett.\ B \textbf{361}, 160 (1995) [arXiv:hep-ph/9507449]. 

\bibitem{Sexton:1996ed} J.~Sexton, A.~Vaccarino and D.~Weingarten, 
Nucl.\ Phys.\ Proc.\ Suppl.\ \textbf{47}, 128 (1996)
[arXiv:hep-lat/9602022]. 

\bibitem{Venugopal:1998fq} E.~P.~Venugopal and B.~R.~Holstein, 
Phys.\ Rev.\ D \textbf{57}, 4397 (1998) [arXiv:hep-ph/9710382]. 
\bibitem{Kawarabayashi:1980dp}
  K.~Kawarabayashi and N.~Ohta,
  Nucl.\ Phys.\ B {\bf 175} (1980) 477.  

\bibitem{Gasser:1984gg} J.~Gasser and H.~Leutwyler, 
Nucl.\ Phys.\ B \textbf{250} (1985) 465; 
R.~Kaiser and H.~Leutwyler, 
Eur.\ Phys.\ J.\ C \textbf{17} (2000) 623 [arXiv:hep-ph/0007101]; 
T.~Feldmann, 
Int.\ J.\ Mod.\ Phys.\ A \textbf{15}, 159 (2000) [arXiv:hep-ph/9907491]. 

\bibitem{Minkowski:2002nf} P.~Minkowski and W.~Ochs, 
Nucl.\ Phys.\ Proc.\ Suppl.\ \textbf{121}, 123 (2003)
[arXiv:hep-ph/0209225]; 
P.~Minkowski and W.~Ochs, 
arXiv:hep-ph/9905250. 

\bibitem{Minkowski:1998mf} P.~Minkowski and W.~Ochs, 
Eur.\ Phys.\ J.\ C \textbf{9}, 283 (1999) [arXiv:hep-ph/9811518]. 

\bibitem{Lebed:1998st} R.~F.~Lebed, 
Czech.\ J.\ Phys.\ \textbf{49} (1999) 1273 [arXiv:nucl-th/9810080]. 

\bibitem{Hatsuda:1994pi} T.~Hatsuda and T.~Kunihiro, 
Phys.\ Rept.\ \textbf{247}, 221 (1994) [arXiv:hep-ph/9401310]. 

\bibitem{Klevansky:qe} S.~P.~Klevansky, 
Rev.\ Mod.\ Phys.\ \textbf{64}, 649 (1992). 

\bibitem{Barberis:2000cd} D.~Barberis \textit{et al.} [WA102 Collaboration], 
Phys.\ Lett.\ B \textbf{479}, 59 (2000) [arXiv:hep-ex/0003033]. 

\bibitem{Amsler:2002ey} C.~Amsler, 
Phys.\ Lett.\ B \textbf{541}, 22 (2002) [arXiv:hep-ph/0206104]. 

\bibitem{Amsler:1997up} C.~Amsler, 
Rev.\ Mod.\ Phys.\ \textbf{70}, 1293 (1998) [arXiv:hep-ex/9708025]. 

\bibitem{Abele:2001pv} A.~Abele \textit{et al.} [CRYSTAL BARREL
Collaboration], 
Eur.\ Phys.\ J.\ C \textbf{21}, 261 (2001). 

\bibitem{Thoma:2003in} U.~Thoma, 
Eur.\ Phys.\ J.\ A \textbf{18}, 135 (2003). 

\bibitem{Bugg:1996ki} D.~V.~Bugg, B.~S.~Zou and A.~V.~Sarantsev, 
Nucl.\ Phys.\ B \textbf{471}, 59 (1996). 

\bibitem{Baker:2003jh} C.~A.~Baker \textit{et al.}, 
Phys.\ Lett.\ B \textbf{563}, 140 (2003). 

\bibitem{Gobbi:1993au} C.~Gobbi, F.~Iachello and D.~Kusnezov, 
Phys.\ Rev.\ D \textbf{50}, 2048 (1994) [arXiv:hep-ph/9310250]. 

\bibitem{Ackleh:1996yt} E.~S.~Ackleh, T.~Barnes and E.~S.~Swanson, 
Phys.\ Rev.\ D \textbf{54}, 6811 (1996) [arXiv:hep-ph/9604355]. 

\bibitem{Groom:in} D.~E.~Groom \textit{et al.} [Particle Data Group
Collaboration], 
Eur.\ Phys.\ J.\ C \textbf{15}, 1 (2000). 

\bibitem{Longacre:1986fh} R.~S.~Longacre \textit{et al.}, 
Phys.\ Lett.\ B \textbf{177}, 223 (1986). 

\bibitem{Li:1990sx} Z.~P.~Li, F.~E.~Close and T.~Barnes, 
Phys.\ Rev.\ D \textbf{43}, 2161 (1991). 
\end{thebibliography}
\end{document}